\documentclass[twocolumn,trackchanges]{aastex61}
\usepackage{amsmath,amstext}
\usepackage[T1]{fontenc}

\received{date}
\revised{date}
\accepted{date}
\submitjournal{ApJL}

\shorttitle{Molecular gas in debris disks around young A-type stars}
\shortauthors{Mo\'or et al.}

\begin{document}

\title{Molecular gas in debris disks around young A-type stars}

\correspondingauthor{Attila Mo\'or}
\email{moor@konkoly.hu}

\author{Attila Mo\'or}
\affil{Konkoly Observatory, Research Centre for Astronomy and
  Earth Sciences, Hungarian Academy of Sciences, Konkoly-Thege
  Mikl\'os \'ut 15-17, 1121 Budapest, Hungary}

\author{Michel Cur\'e} 
\affiliation{Instituto de F\'\i{}sica y
  Astronom\'\i{}a, Universidad de Valpara\'\i{}so, Valpara\'\i{}so,
  Chile}

\author{\'Agnes K\'osp\'al} \affiliation{Konkoly Observatory, Research
  Centre for Astronomy and Earth Sciences, Hungarian Academy of
  Sciences, Konkoly-Thege Mikl\'os \'ut 15-17, 1121 Budapest, Hungary}
  \affiliation{Max Planck Institute for
    Astronomy, K\"onigstuhl 17, D-69117 Heidelberg, Germany}

\author{P\'eter \'Abrah\'am} 
\affiliation{Konkoly Observatory,
  Research Centre for Astronomy and Earth Sciences, Hungarian Academy
  of Sciences, Konkoly-Thege Mikl\'os \'ut 15-17, 1121 Budapest,
  Hungary}

\author{Timea Csengeri} \affiliation{Max Planck Institute for Radio
  Astronomy, Auf dem H\"ugel 69, 53121 Bonn, Germany}

\author{Carlos Eiroa} \affiliation{Dpto. F\'\i sica Te\'orica, Universidad Aut\'onoma de Madrid, 
Cantoblanco, 28049, Madrid, Spain}

\author{Diah Gunawan} \affiliation{Instituto de F\'\i{}sica y
  Astronom\'\i{}a, Universidad de Valpara\'\i{}so, Valpara\'\i{}so,
  Chile}

\author{Thomas Henning} \affiliation{Max Planck Institute for
  Astronomy, K\"onigstuhl 17, D-69117 Heidelberg, Germany}

\author{A. Meredith Hughes} \affiliation{Department of Astronomy, Van
  Vleck Observatory, Wesleyan University, 96 Foss Hill Drive,
  Middletown, CT 06459, USA}

\author{Attila Juh\'asz} \affiliation{Institute of Astronomy,
  University of Cambridge, Madingley Road, Cambridge CB3 0HA, UK}
submission/
\author{Nicole Pawellek} \affiliation{Max Planck Institute for
  Astronomy, K\"onigstuhl 17, D-69117 Heidelberg, Germany}

\author{Mark Wyatt} \affiliation{Institute of Astronomy, University of
  Cambridge, Madingley Road, Cambridge CB3 0HA, UK}


\begin{abstract}
According to the current paradigm of circumstellar disk evolution,
gas-rich primordial disks evolve into gas-poor debris disks composed
of second-generation dust.
To explore the transition between these phases,
we searched for $^{12}$CO, $^{13}$CO, and C$^{18}$O emission in seven 
{dust-rich} debris disks around young A-type stars, using ALMA in Band 6.
We discovered molecular gas in three debris disks. 
In all these disks, the $^{12}$CO line was optically thick, 
highlighting the importance of less abundant molecules in reliable mass
estimates.
Supplementing our target list by literature data,
we compiled a volume-limited sample of {dust-rich} debris disks 
{around young A-type stars within 150\,pc.} 
{We obtained a CO detection rate of 11/16 above a $^{12}$CO J=2--1
line luminosity threshold of $\sim$1.4$\times$10$^4$\,Jy~km~s$^{-1}$pc$^2$ 
in the sample. This high incidence implies that the presence of CO gas in 
bright debris disks around young A-type stars is likely more the rule than the exception.}
{Interestingly, dust-rich debris disks around young FG-type stars exhibit,
with the same detectability threshold as for A-type stars, significantly lower 
gas incidence.}
While the transition from protoplanetary to debris phase is associated with
a drop of dust content, our results exhibit a large spread in the CO mass in our debris sample,
with peak {values comparable to those} in protoplanetary Herbig Ae disks.
In the {particularly} CO-rich debris systems the gas may have primordial origin, 
characteristic of a hybrid disk.
\end{abstract}



\section{Introduction}
\label{sec:intro}

During their early evolution, newborn stars
are surrounded by massive gas-rich circumstellar disks.
Most these primordial disks dissipate by the age of 10\,Myr \citep[e.g.,][]{alexander2014}, 
partly because their material is incorporated to planetesimals and planets.
Later, the collisional erosion of the
planetesimals may produce fresh dust \citep{wyatt2008}.
The tenuous debris disk, formed by these second-generation dust
particles, may accompany the star almost throughout its life. The
transformation from primordial to debris disk is perhaps the most
radical change during a disk's lifetime, whose details are still
little understood \citep{wyatt2015}. The evolution of the dust
component is relatively well known, because infrared and submillimeter
continuum observations outline how the dust mass decreases with time
\citep{wyatt2008}. Due to the rarity of gas detections in debris
disks, we know significantly less about the gas component. Mature
debris disks are expected to have low gas-to-dust ratio, because
processes like collisions, sublimation, and photodesorption from icy
grains or planetesimals can produce only a small amount of
secondary gas and moreover the most detectable species like CO 
photodissociate rapidly in the interstellar radiation field \citep{matthews2014}.

Recent surveys with single dish radio
telescopes and the ALMA interferometer identified a growing
population of debris disks containing {detectable amounts of} molecular gas.
{Detecting rotational transitions of CO molecules, 
up to now twelve CO-bearing debris disks have been discovered: 
49\,Cet \citep{zuckerman1995}, HD\,21997 \citep{moor2011}, $\beta$\,Pic \citep{dent2014},   
HD\,131835 \citep{moor2015a},  HD\,181327 \citep{marino2016}, HD\,110058, HD\,138813, HD\,146897, 
HD\,156623 \citep{lieman-sifry2016}, HD\,32297 \citep{greaves2016}, $\eta$\,Crv \citep{marino2017}, 
and Fomalhaut \citep{matra2017b}. These disks share several distinctive physical
characteristics: most of them are young ($<$50\,Myr), 
surround A-type stars, exhibit high fractional luminosity, and have a dust component that is
relatively cold ($<$140\,K), resembling the Kuiper belt rather than the asteroid
belt in Solar System terminology. The gas component may have a secondary origin, {being produced 
from icy material within the planetesimal belt.}However, considering the youth of the systems, 
it may also be the remnant of the protoplanetary disk ({\it primordial origin}).
There is now proof for secondary gas
in four systems: $\beta$\,Pic, \citep{matra2017}, Fomalhaut \citep{matra2017b}, HD\,181327 \citep{marino2016}, and $\eta$\,Crv \citep{marino2017}.   
The large amount of gas in HD~21997 and HD~131835
{is difficult to explain within the framework of current secondary 
gas production models \citep{kospal2013,kral2017}, raising the possibility that 
the gas is primordial while the dust is secondary,}
forming a {\it hybrid disk} (\citealt{kospal2013}, Mo\'or et al.~in
prep.).} 

We have only a few constraints on how circumstellar disks reach the gas-poor
phase, how closely the gas evolution is coupled to the disappearance
of the primordial dust, when the secondary gas production starts, and how long 
the disk can retain the primordial gas. Motivated by these
open questions and to explore the conditions under which a disk could
keep its primordial gas component longer, we initiated a systematic
investigation of molecular gas in debris disks. We focus on young
A-type stars, and determine the incidence and physical parameters of
their molecular gas component. Here we report on new ALMA observations of
seven disks, which were selected to complete a volume limited
sample of {dust-rich} debris disks around young A-type stars within
150\,pc. We targeted all three main CO isotopologues, 
since $^{12}$CO alone does not provide reliable gas mass
estimates if optically thick.


\section{Sample selection}
\label{sec:targetsel}

In order to obtain a complete census of all debris systems within 150
pc which are similar to the known CO-bearing debris disks, we adopted
the following criteria: (1) A-type host star; (2) $5\times10^{-4} < L_{\rm
  disk} / L_{\rm bol} < 10^{-2}$, thus the fractional
luminosity is higher than the lowest value {(in HD\,21997) found in any CO-bearing 
debris disk younger than 50 Myr,} but lower than typical values in protoplanetary disks;
(3) dust temperature $<$140\,K, to ensure that -- similarly to all
previous detections -- the disks have a cold dust component; (4)
$\ge$70\,{\micron} detection with Spitzer/MIPS or Herschel/PACS; (5)
age between 10 and 50\,Myr.  We
searched the literature \citep{ballering2013,chen2014,melis2013,
moor2011} and found 17 systems within 150\,pc which satisfied these
criteria.
{The age estimates for all systems are well established, 
being based on membership in young moving groups or associations,
except one target, HD\,32297.}  
We excluded from this sample those 10 systems where
sensitive ALMA observations were already available. The remaining
seven disks formed the target list of our dedicated ALMA survey.
Remarkably, all these objects belong to the Sco-Cen association.
Table~\ref{tab:props} gives the main parameters for all 17 sources,
including our present targets.


\begin{deluxetable*}{lccccc|cCC|cCC}
\tablecaption{Stellar and disk parameters \label{tab:props}}
\tablecolumns{12}
\tablewidth{700pt}
\tabletypesize{\scriptsize}
\tablehead{
\colhead{Name} & \colhead{Spt.} & \colhead{Dist.} & \colhead{Lum.} & \colhead{Group} &
\colhead{Age} & \colhead{$T_{\rm dust}$} & \colhead{$L_{\rm disk}/L_{\rm bol}$} & \colhead{$M_{\rm dust}$} & 
\colhead{CO} & \colhead{$S_{\rm ^{12}CO}$} & \colhead{$M_{\rm CO}$}   \\
\colhead{} & \colhead{} & \colhead{(pc)} & \colhead{($L_*$)} & \colhead{} &
\colhead{(Myr)} & \colhead{(K)} & \colhead{} & \colhead{($M_{\oplus}$)} & 
\colhead{} & \colhead{(Jy~km~s$^{-1}$)} & \colhead{($M_{\oplus}$)} 
} 
\colnumbers
\startdata
\cutinhead{Current sample}
  HD\,98363 &	   A2V &  123.6 &  10.5 &   LCC (7) &  15 (4) &  295/112 &  9.2$\times$10$^{-4}$ (2)       & 6.8$\times$10$^{-3}$ (8)      &    N & $<$0.036 (8)      &\nodata \\
 HD\,109832 &	   A9V &  111.9 &   5.3 &   LCC (7) &  15 (4) &   186/92 &  7.6$\times$10$^{-4}$ (1)       & $<$6.7$\times$10$^{-3}$ (8)   &    N & $<$0.033 (8)      &\nodata \\
 HD\,121191 &	A5IV/V &  135.9 &   8.2 &   LCC/UCL (4) & 15--16 (4) &  555/118 & { 4.7$\times$10$^{-3}$} (9)& 9.5$\times$10$^{-3}$ (8)      &    Y & 0.23$\pm$0.04 (8) & { 2.7$\times$10$^{-3}$} \\
 HD\,121617 &	   A1V &  128.2 &  17.0 &   UCL (2) &  16 (4) &      105 &  4.8$\times$10$^{-3}$ (4)       & 1.4$\times$10$^{-1}$ (8)      &    Y & 1.27$\pm$0.13 (8) & { 1.8$\times$10$^{-2}$} \\
 HD\,131488 &	   A1V &  147.7 &  13.1 &   UCL (4) &  16 (4) &   570/94 &  5.5$\times$10$^{-3}$ (9)       & 3.2$\times$10$^{-1}$ (8)      &    Y & 0.78$\pm$0.08 (8) & { 8.9$\times$10$^{-2}$} \\
 HD\,143675 &	A5IV/V &  113.4 &   7.1 &   UCL (7) &  16 (4) &  374/127 &  5.9$\times$10$^{-4}$ (2)       & $<$1.1$\times$10$^{-2}$ (8)   &    N & $<$0.078 (8)      &\nodata \\
 HD\,145880 &	 B9.5V &  127.9 &  32.8 &   UCL (7) &  16 (4) &   196/70 &  1.1$\times$10$^{-3}$ (1)       & $<$2.6$\times$10$^{-2}$ (8)   &    N & $<$0.071 (8)      &\nodata \\
 \cutinhead{Other systems}   
  49\,Cet &	   A1V &   59.4 &  16.4 &   ARG (8) &  40 (5) &   155/56 &  1.1$\times$10$^{-3}$ (5)       & 1.7$\times$10$^{-1}$ (2)      &    Y & 2.00$\pm$0.30 (3) & $\geq$1.9$\times$10$^{-4}$ \\
  HD\,21997 &	A3IV/V &   71.9 &  11.2 &   COL (5) &  42 (1) &       61 &  5.7$\times$10$^{-4}$ (5)       & 1.1$\times$10$^{-1}$ (7)      &    Y & 2.17$\pm$0.23 (4) & 6.0$\times$10$^{-2}$ \\
  HD\,32297 &	   A6V &  112.4 &   6.2 &     -     &  $<$30 (2) &   292/88 &  4.4$\times$10$^{-3}$ (2)       & 2.0$\times$10$^{-1}$ (4)      &    Y & 5.11$\pm$0.49 (1) & $\geq$1.5$\times$10$^{-3}$ \\
  $\beta$\,Pic &   A6V &   19.4 &   8.7 &  BPMG (1) &  23 (3) &       85 &  2.6$\times$10$^{-3}$ (3,7)     & 4.2$\times$10$^{-2}$ (1)      &    Y & 4.55$\pm$0.52 (6) & 3.4$\times$10$^{-5}$ \\
  HD\,95086 &	 A8III &   90.4 &   7.0 &   LCC (7) &  15 (4) &   184/54 &  1.7$\times$10$^{-3}$ (5)       & 2.2$\times$10$^{-1}$ (6)      &    N & $<$0.014 (7)      & \nodata \\
   HR\,4796 &	   A0V &   72.8 &  23.4 &   TWA (6) &  10 (5) &      108 &  4.6$\times$10$^{-3}$ (8)       & 1.4$\times$10$^{-1}$ (5)      &    N & $<$6.15 (2)       & \nodata \\
 HD\,110058 &	   A0V &  107.4 &   5.9 &   LCC (7) &  15 (4) &  499/112 &  1.4$\times$10$^{-3}$ (2)       & 2.9$\times$10$^{-2}$ (3)      &    Y & 0.09$\pm$0.02 (5) & $\geq$2.5$\times$10$^{-5}$ \\
 HD\,131835 &	  A2IV &  122.7 &   9.2 &   UCL (7) &  16 (4) &   176/71 &  3.0$\times$10$^{-3}$ (6)       & 2.5$\times$10$^{-1}$ (3)      &    Y & 0.80$\pm$0.04 (5) & { 3.2$\times$10$^{-2}$} \\
 HD\,138813 &	   A0V &  150.8 &  24.5 &    US (7) &  10 (4) &   194/94 &  9.0$\times$10$^{-4}$ (2)       & 1.2$\times$10$^{-1}$ (3)      &    Y & 1.41$\pm$0.08 (5) & $\geq$7.5$\times$10$^{-4}$ \\
 HD\,156623 &	   A0V &  118.3 &  14.8 &   UCL (3) &  16 (4) &  605/123 &  5.0$\times$10$^{-3}$          & 3.2$\times$10$^{-2}$ (3)      &    Y & 1.18$\pm$0.04 (5)  & $\geq$3.9$\times$10$^{-4}$ \\
\enddata
\tablecomments{
Column (1): Target name. 
Column (2): Spectral type (from SIMBAD, except for HD\,32297, see \citealt{rodigas2014}). 
Column (3): Distance (from {\sl Hipparcos} parallax
when available \citep{vanleeuwen2007}, except for HD\,121617 and HD\,121191, see 
\citealt{tgas}, and for HD\,131488, whose kinematic distance was derived based on its membership in UCL).
Column (4): Luminosity.
Column (5): Moving group membership. ARG: Argus moving group; 
	BPMG: $\beta$\,Pic moving group; COL: Columba moving group; 
	   LCC: Lower Centaurus Crux association; UCL: Upper Centaurus
	Lupus association; US: Upper Scorpius association. 
	{References for membership assignments:
       (1) \citet{barrado1999}; (2) \citet{hoogerwerf2000}; (3) \citet{lieman-sifry2016}; 
       (4) \citet{melis2013}; (5) \citet{moor2006}; (6) \citet{webb1999}; (7) \citet{dezeeuw1999}; 
       (8) \citet{zuckerman2012}.}
Column (6): Stellar age. {References for age estimates 
(the age of the corresponding group in case of group members): (1) \citet{bell2015}; 
(2) \citet{kalas2005}; (3) \citet{mamajek2014}; (4) \citet{pecaut2016}; (5) \citet{torres2008}.}
Column (7): Dust temperature. For references see Col.~(8). 
Column (8): Fractional luminosity of the disk. References:
(1) \citet{ballering2013}; (2) \citet{chen2014}; (3) \citet{dent2014}; (4) \citet{moor2011};
(5) \citet{moor2015a}; (6) \citet{moor2015b}; (7) \citet{rhee2007}; (8) \citet{rm2014}; (9) \citet{vican2016}. 
For HD\,156623, the characteristic dust temperatures and the fractional 
luminosity were derived by fitting a two-component modified blackbody to the excess SED.
Column (9): Dust mass.  
 References for the utilized (sub)millimeter observations: 
 (1) \citet{dent2014}; (2) \citet{hughes2017}; (3) \citet{lieman-sifry2016}; (4) \citet{meeus2012};
(5) \citet{sheret2004}; (6) Su et al. in prep.; (7) \citet{williams2006} ; (8) this work.
Column (10): CO detection status.
Column (11): Integrated line fluxes {or 3$\sigma$ upper limits} for $^{12}$CO $J$=2$-$1 ($J$=3$-$2 for HR\,4796). 
References: (1) \citet{greaves2016}; (2) \citet{hales2014}; (3) \citet{hughes2008}; (4) \citet{kospal2013};
(5) \citet{lieman-sifry2016}; (6) \citet{matra2017}; (7) Su et al. in prep.; (8) this work.
Column (12): Mass of CO gas (for details see Sect.~\ref{sec:res}). 
}
\end{deluxetable*}


\section{Observations}
\label{sec:obs}
Our targets were observed with ALMA in Band\,6 
(project 2015.1.01243.S, PI: M.~Cur\'e). Table~\ref{tab:log}
shows the log of observations. We used identical correlator setup
in each case. Two spectral windows centered at 217 and 233.5\,GHz
were dedicated to continuum measurements, providing 1875\,MHz
bandwidth individually. The other two spectral windows were tuned
to cover the $^{12}$CO, $^{13}$CO, and C$^{18}$O J=2--1 lines. {Observations 
of the $^{12}$CO(2-1) transition were performed 
using a window centered at 230.743\,GHz with a bandwidth of 468.75\,MHz,
while the isotopologue lines were measured together in a window centered 
at 219.492\,GHz with a bandwidth of 1875\,MHz. The spectral resolutions
in the two windows were 244\,kHz (0.32\,kms$^{-1}$) and 976\,kHz (1.33\,kms$^{-1}$),
respectively.}Calibration and imaging were done using the standard ALMA reduction
tool {\sl Common {Astronomy} Software Applications}
\citep[CASA v4.5;][]{mcmullin2007}. To extract the continuum image and
spectral cubes of different lines from the calibrated visibilities, we
used the CASA task \texttt{clean} with Briggs weighting using a
robustness parameter of 0.5. For the line data, we first fitted a
continuum to the line-free channels and subtracted it in the {\it uv}
space using the \texttt{uvcontsub} task. The continuum image was
compiled by concatenating the dedicated continuum spectral windows and
line free channels of the other two spectral windows. Beam sizes,
position angles, and rms noises are also summarized in
Table~\ref{tab:log}.


\begin{deluxetable*}{lccccccc}
\tablecaption{Observational data\label{tab:log}}  
\tablecolumns{8}
\tabletypesize{\scriptsize}
\tablewidth{0pt}
\tablehead{
\colhead{Parameters} & 
\colhead{HD\,98363}  &
\colhead{HD\,109832} &
\colhead{HD\,121191} &
\colhead{HD\,121617} &
\colhead{HD\,131488} &
\colhead{HD\,143675} &
\colhead{HD\,145880}
}
\startdata
\cutinhead{{ Observational and imaging parameters}}
Obs. date             & 2016Mar 22 & 2016 Mar 22 & 2016 May 23 & 2016 May 23 & 2016 May 08 & 2016 May 16 & 2016 May 16 \\
Number of antennas    & 37  & 37 & 35 & 35 & 41 & 39 & 39 \\
Baseline lengths (m)  & 15.3--460 & 15.3--460 & 16.7--640 & 16.7--640 & 15.1--640 & 16.5--640 & 16.5--640 \\
\cutinhead{Continuum imaging}
Beam size ({\arcsec}) & 0$\farcs$70$\times$0$\farcs$82 & 0$\farcs$71$\times$0$\farcs$86 & 0$\farcs$51$\times$0$\farcs$59 & 0$\farcs$49$\times$0$\farcs$57 & 0$\farcs$53$\times$0$\farcs$55 & 0$\farcs$48$\times$0$\farcs$62 & 0$\farcs$48$\times$0$\farcs$64 \\ 
Beam PA ({\degr}) & 32$\fdg$5 & $-$6$\fdg$7 & $-$50$\fdg$3 & $-$64$\fdg$1 & $-$34$\fdg$6 & $-$51$\fdg$9 & $-$51$\fdg$3 \\
rms ($\mu$Jy~beam$^{-1}$) & 26 & 25 & 40 & 42 & 29 & 43 & 43 \\
\cutinhead{$^{12}$CO (2-1) imaging}
Beam size ({\arcsec}) & 0$\farcs$71$\times$0$\farcs$84 & 0$\farcs$71$\times$0$\farcs$88 & 0$\farcs$52$\times$0$\farcs$60 & 0$\farcs$51$\times$0$\farcs$58 & 0$\farcs$54$\times$0$\farcs$56  & 0$\farcs$48$\times$0$\farcs$62 & 0$\farcs$48$\times$0$\farcs$65 \\ 
Beam PA ({\degr}) & $+$29$\fdg$7  & $-$7$\fdg$1 & $-$49$\fdg$5 & $-$63$\fdg$4 & $-$16$\fdg$4 & $-$50$\fdg$9 & $-$49$\fdg$8 \\
rms (mJy~beam$^{-1}$~chan.$^{-1}$) & 4.3 & 4.2 & 6.2 & 6.1 & 4.5 & 6.4 & 6.3 \\
\cutinhead{$^{13}$CO (2-1) imaging}
Beam size ({\arcsec}) & 0$\farcs$73$\times$0$\farcs$87 & 0$\farcs$74$\times$0$\farcs$91 & 0$\farcs$54$\times$0$\farcs$61 & 0$\farcs$52$\times$0$\farcs$60 & 0$\farcs$56$\times$0$\farcs$57 & 0$\farcs$50$\times$0$\farcs$65 & 0$\farcs$50$\times$0$\farcs$67 \\ 
Beam PA ({\degr}) & $+$31$\fdg$6 &$-$5$\fdg$5  & $-$52$\fdg$2 & $-$65$\fdg$0 & $-$25$\fdg$7 & $-$52$\fdg$9 & $-$51$\fdg$5 \\
rms (mJy~beam$^{-1}$~chan.$^{-1}$) & 2.0  & 2.0 & 3.0 & 2.9 & 2.0 & 2.9 & 2.7 \\
\cutinhead{C$^{18}$O (2-1) imaging}
Beam size ({\arcsec}) & 0$\farcs$74$\times$0$\farcs$88 & 0$\farcs$75$\times$0$\farcs$92 & 0$\farcs$54$\times$0$\farcs$63 & 0$\farcs$53$\times$0$\farcs$62 & 0$\farcs$57$\times$0$\farcs$58 & 0$\farcs$51$\times$0$\farcs$66 & 0$\farcs$51$\times$0$\farcs$68 \\ 
Beam PA ({\degr}) & $+$30$\fdg$5 & $-$5$\fdg$5 & $-$49$\fdg$1 & $-$63$\fdg$5 & $-$21$\fdg$4 & $-$52$\fdg$5 & $-$50$\fdg$7 \\
rms (mJy~beam$^{-1}$~chan.$^{-1}$) & 1.6 & 1.6 & 2.3 & 2.1 & 1.6 & 2.3 & 2.3 \\
\cutinhead{{ Measured continuum flux densities and CO integrated line fluxes}}
$F_\nu$ at 1.33\,mm (mJy)           & 0.107$\pm$0.027 & $<$0.105 & 0.130$\pm$0.029 & 1.86$\pm$0.29   & 2.91$\pm$0.31     & $<$0.228 & $<$0.234 \\  
$S_{\rm ^{12}CO}$  (Jy~km~s$^{-1}$) & $<$0.036        & $<$0.036 & 0.231$\pm$0.044 & 1.27$\pm$0.13   & 0.780$\pm$0.084   & $<$0.078 & $<$0.071 \\  
$S_{\rm ^{13}CO}$  (Jy~km~s$^{-1}$) & $<$0.036        & $<$0.034 & 0.071$\pm$0.022 & 0.529$\pm$0.065 & 0.490$\pm$0.054   & $<$0.070 & $<$0.068 \\  
$S_{\rm C^{18}O}$  (Jy~km~s$^{-1}$) & $<$0.026        & $<$0.025 & $<$0.050        & 0.079$\pm$0.023 & 0.266$\pm$0.033   & $<$0.061 & $<$0.060 \\
\enddata
\end{deluxetable*}


\section{Results}
\label{sec:res}

\begin{figure*}
\begin{center}
\includegraphics[angle=0,scale=.327]{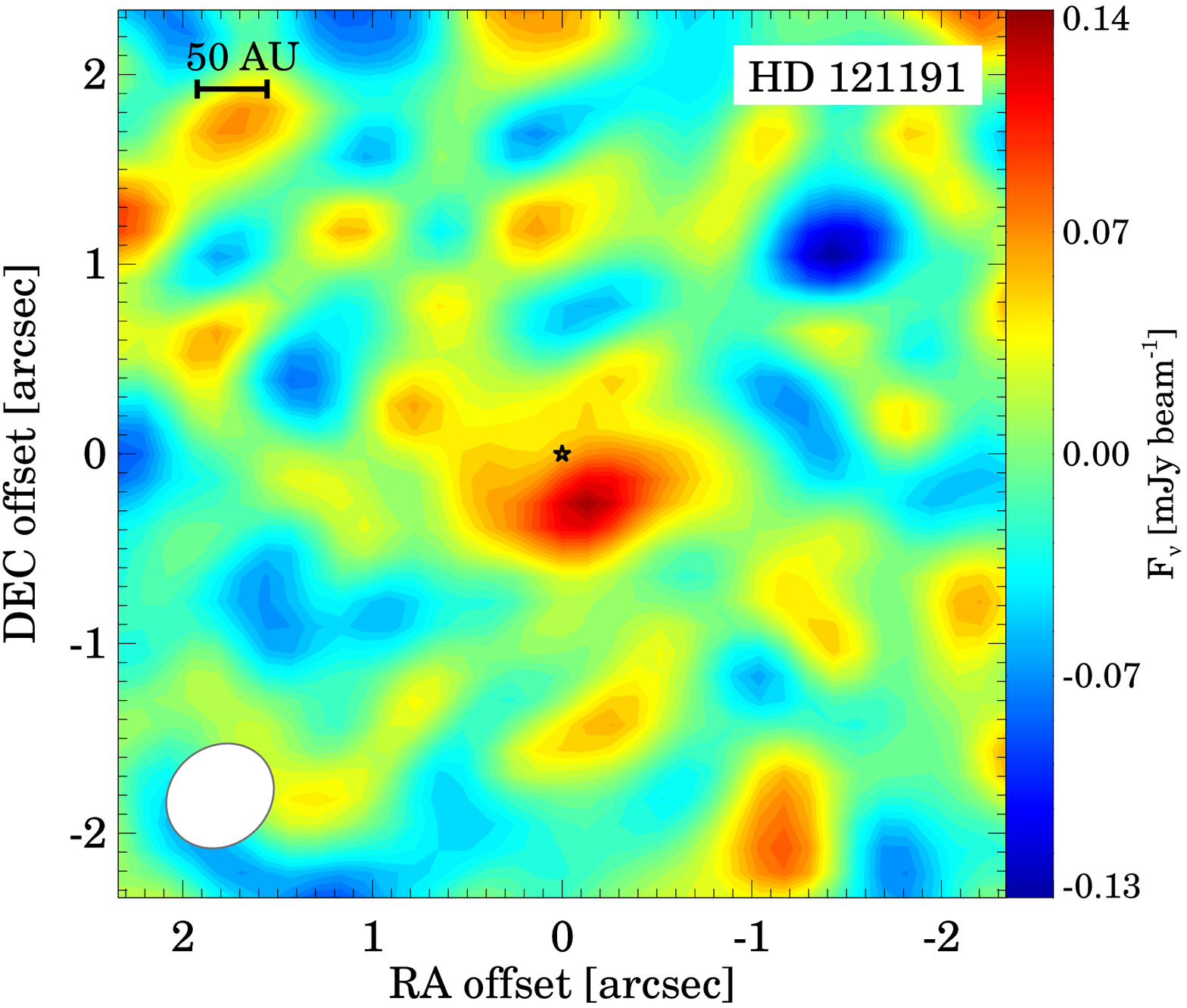}
\includegraphics[angle=0,scale=.327]{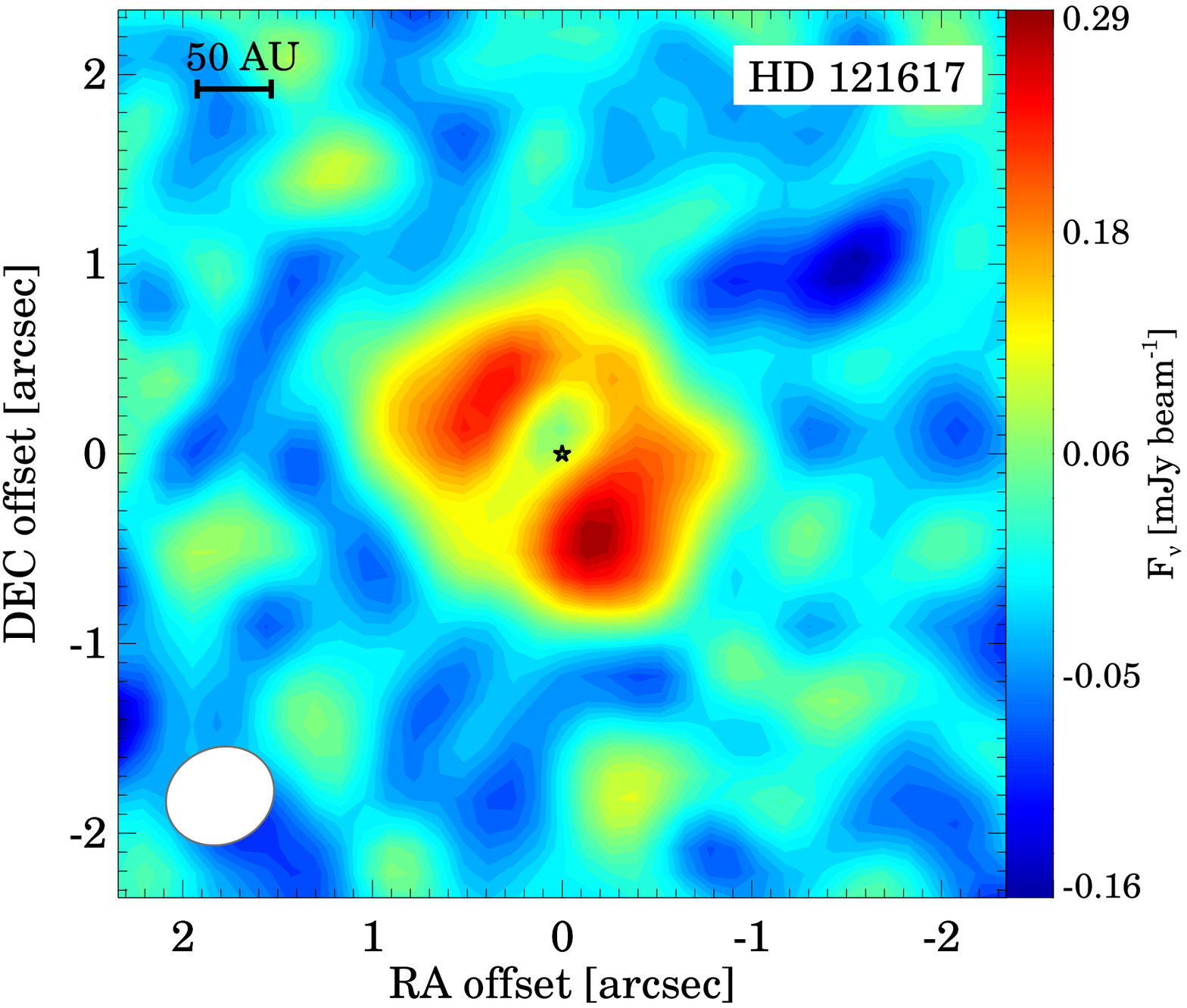}
\includegraphics[angle=0,scale=.327]{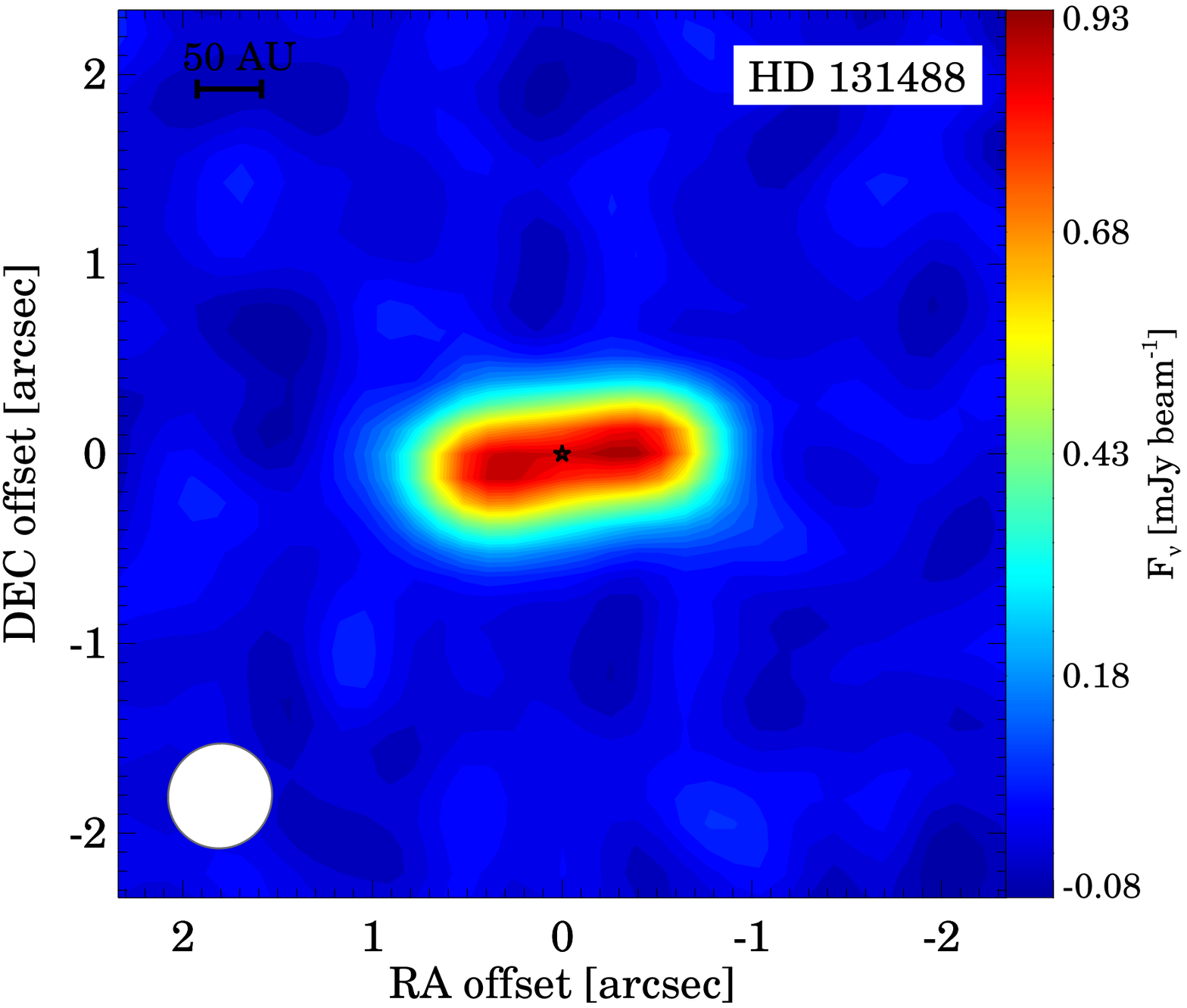}
\includegraphics[angle=0,scale=.24]{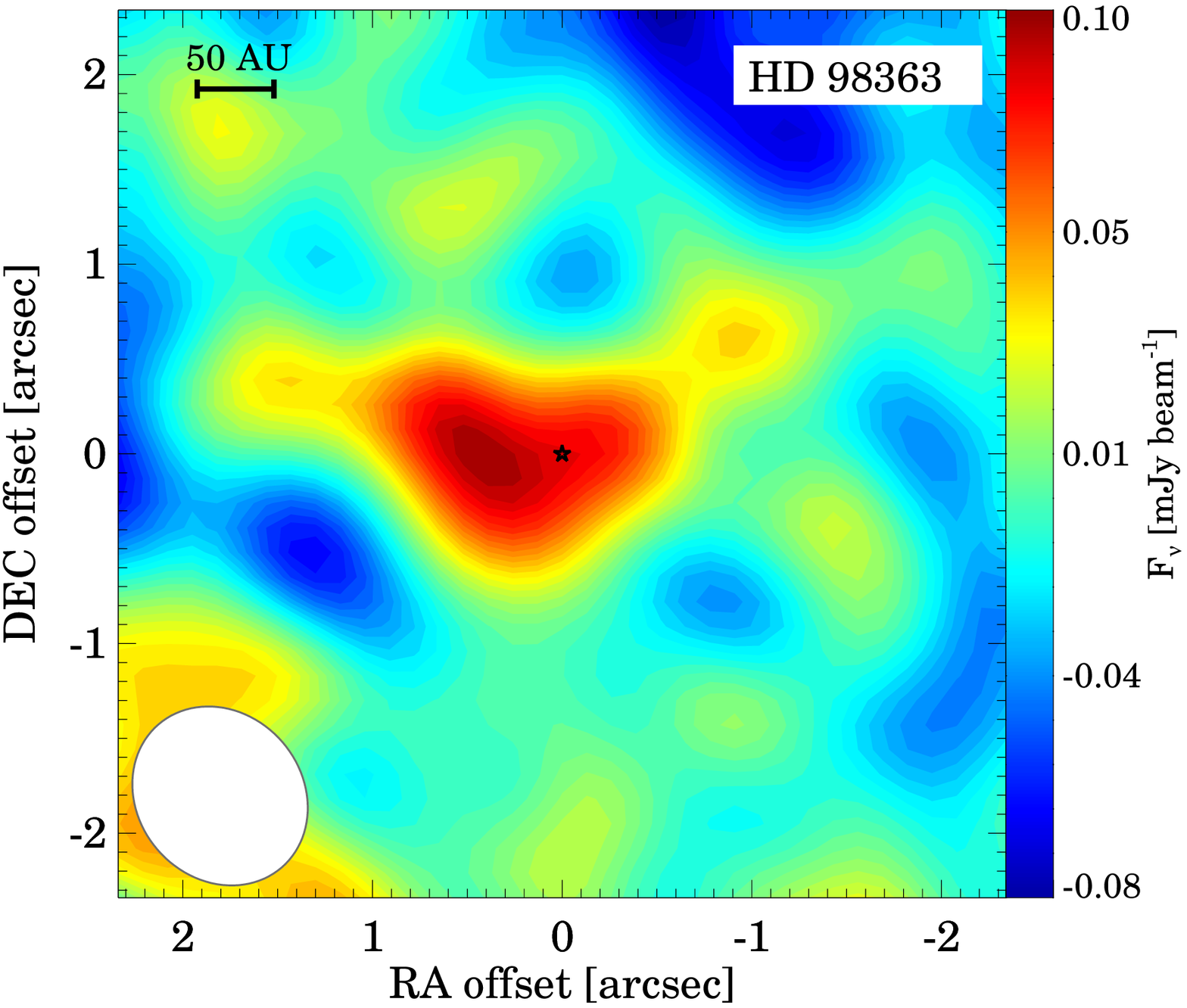}
\includegraphics[angle=0,scale=.24]{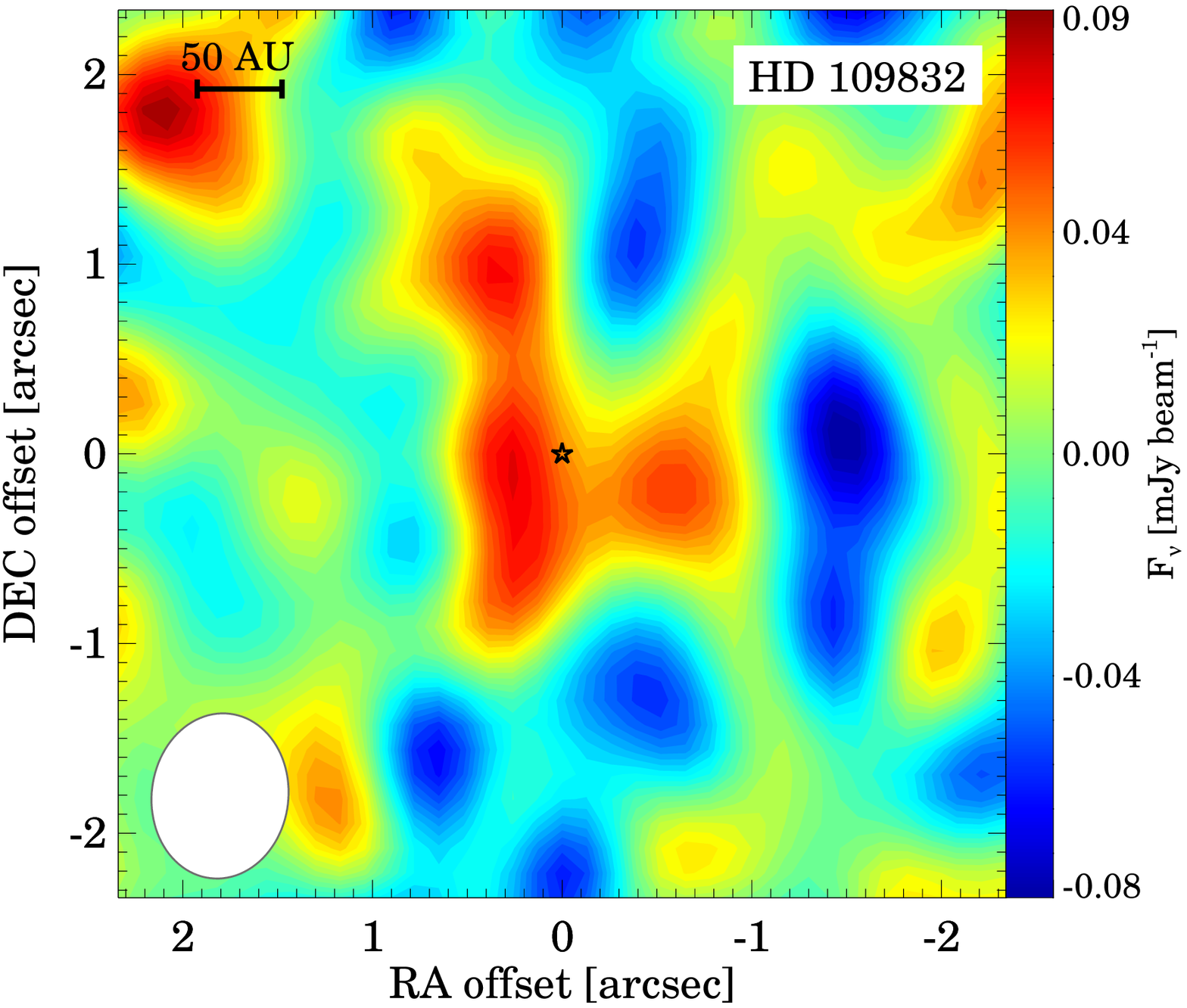}
\includegraphics[angle=0,scale=.24]{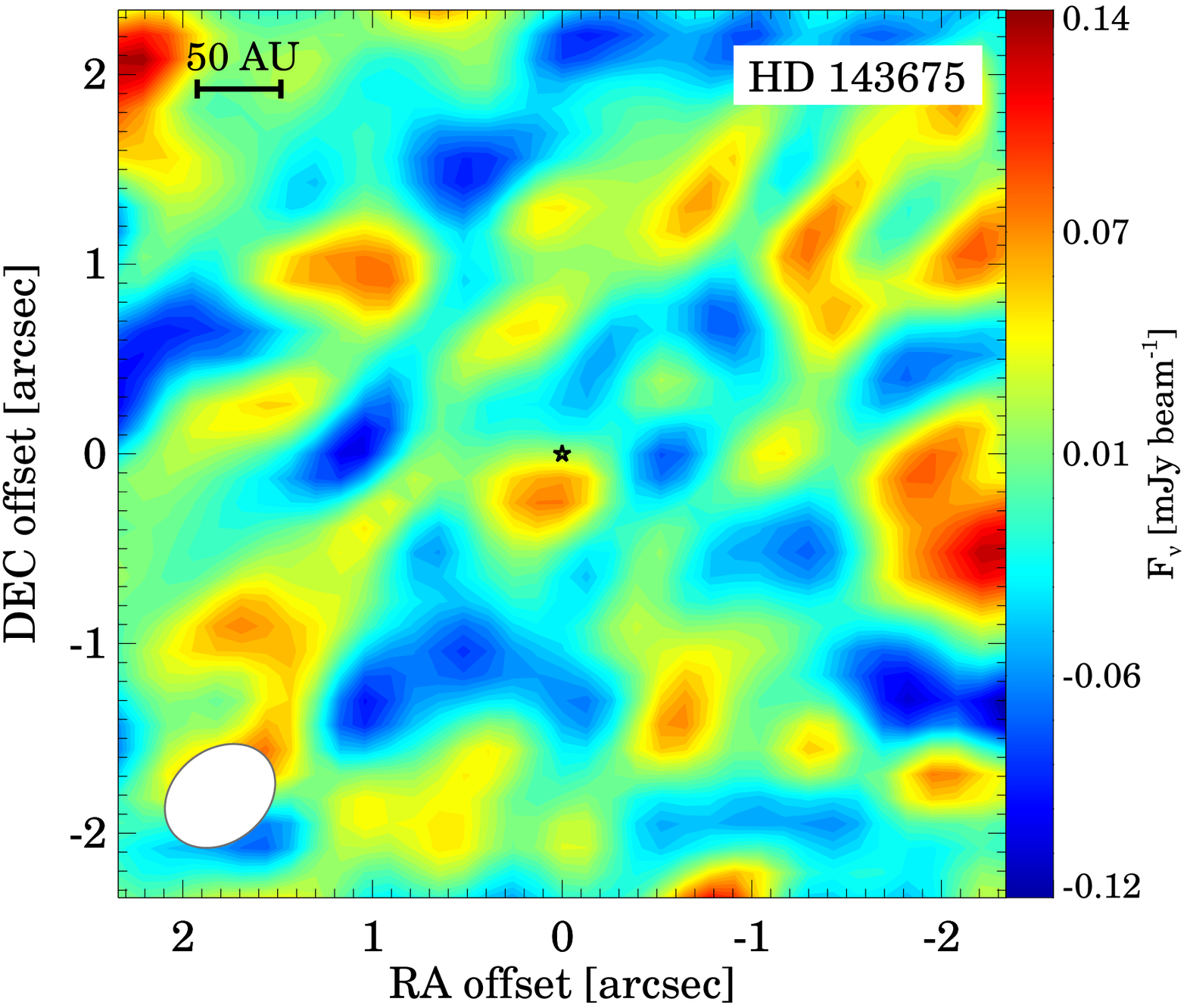}
\includegraphics[angle=0,scale=.24]{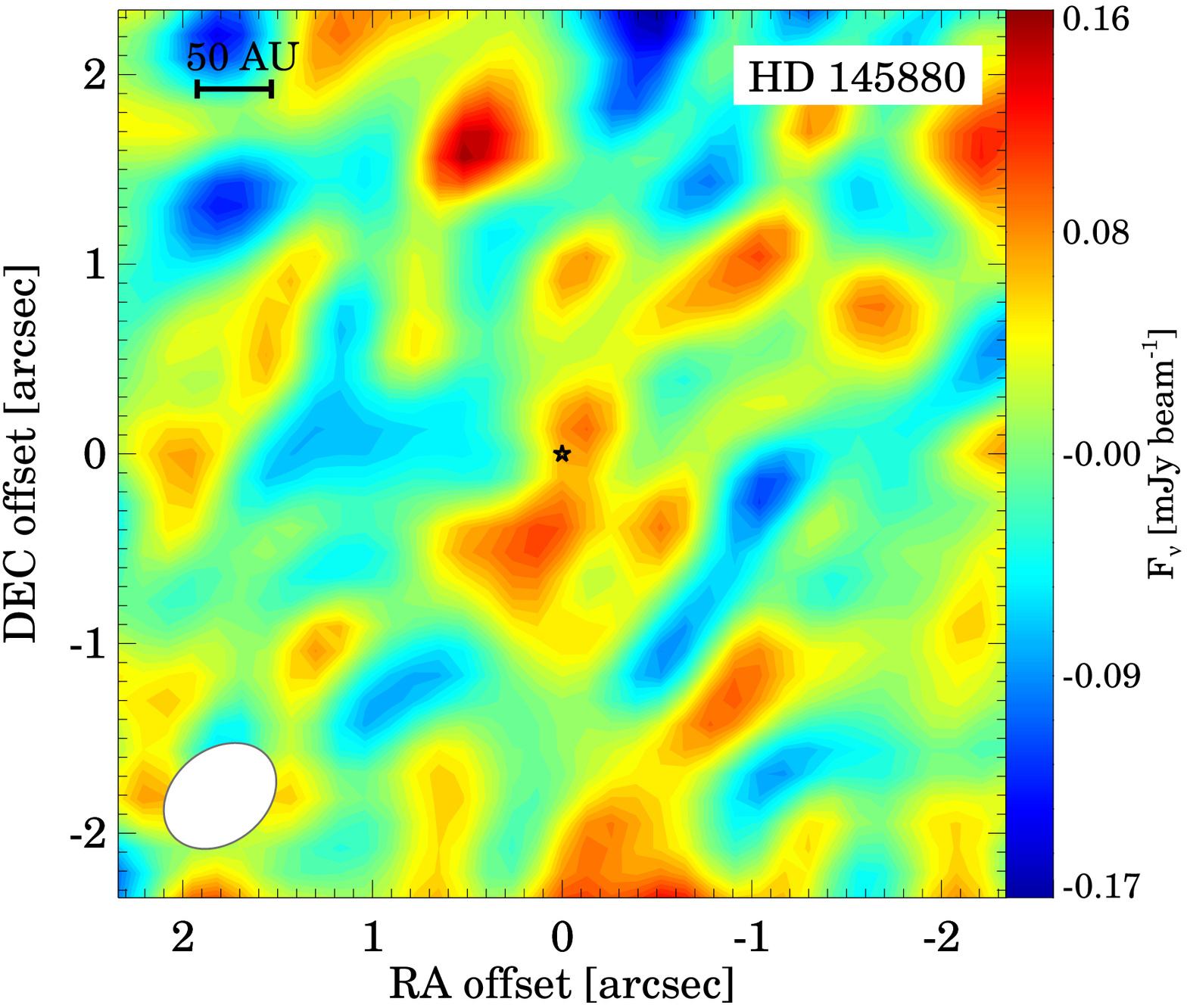}
\end{center}
\caption{ALMA continuum images at 1.33\,mm of our seven targets. Upper panels 
show the maps of CO-bearing disks.
    \label{fig:continuum}}
\end{figure*}

\paragraph{Continuum data.} Fig.~\ref{fig:continuum} presents the
1.3\,mm continuum images of our targets. For HD~121191, HD~121617,
HD~131488, and HD~98363, we detected significant (peak $>3\,\sigma$)
signal towards or close to the position of the central star. For
HD~109832, the {peak} emission close to the stellar position is only
2.8\,$\sigma$. The remaining two objects were undetected. 
Continuum flux densities of HD\,98363 and HD\,121191 were determined 
by fitting a point source to their visibility data using the CASA \texttt{uvmodelfit} task. To derive the integrated flux density 
of the bright extended disks around HD\,121617 and HD\,131488 we used 
the \texttt{uvmultifit} task \citep{marti-vidal2014} to fit a Gaussian ring model to the visibilities. These models also provide structural information 
on these disks (see Sect.~\ref{discovery}).
We determined upper limits for the
non-detected sources by measuring the flux at random positions in the
images in an aperture radius corresponding to 100\,au. The measured
fluxes and 3$\sigma$ upper limits are listed in Tab.~\ref{tab:log}, along
with uncertainties containing 10\% absolute calibration error added
quadratically. 

Assuming optically thin emission, the measured flux
densities were converted to dust masses {using 
\begin{equation}
M_{d} = \frac{{F_{\nu}} d^2}{ B_{\nu}(T_{\rm dust}) \kappa_{\nu}},
\end{equation}
where $d$ is the distance, $B_{\nu} (T_{\rm dust})$ is the Planck
function at 1.3\,mm for a characteristic dust temperature $T_{\rm
  dust}$ taken from Tab.~\ref{tab:props}, while
$\kappa_{\nu}$ is the opacity at 1.3\,mm for which a value of
2.3\,cm$^2$\,g$^{-1}$ was adopted \citep{andrews2013}.}

{For consistency, the dust masses of the other ten disks 
from Table~\ref{tab:props} were also computed using 
submillimeter/millimeter data collected from the literature and adopting 
an opacity value of 
$\kappa_{\nu} = 2.3\left( \frac{\nu}{{\rm 230\,GHz}} \right)^{0.7}$\,cm$^2$\,g$^{-1}$.
Taking into account uncertainties in measured flux densities, distances, 
and dust temperature values, the formal uncertainty of our dust mass estimates are 
between 11 and 30\%. However, there are other factors that can influence 
the estimates. The emitting dust grains were characterized by a single temperature, that 
was derived from the analysis of the SED and therefore reflects the temperature of the 
smallest grains and not of those large ones that we observe at millimeter wavelengths.  
The latter grains are probably colder, thus, by using the SED-based 
$T_{\rm dust}$, we may underestimate the dust mass. 
Even more serious uncertainty can be related to the adopted $\kappa_{\nu}$ value.
Depending on the chemical compositions, shape, and size distribution of dust grains, 
the millimeter opacity can vary significantly \citep{miyake1993,ossenkopf1994}, thereby introducing an 
uncertainty of a factor of several in the calculation.
The obtained dust masses are listed in Table~\ref{tab:props}.
}

\begin{figure*}
\begin{center}
\includegraphics[angle=0,scale=.3]{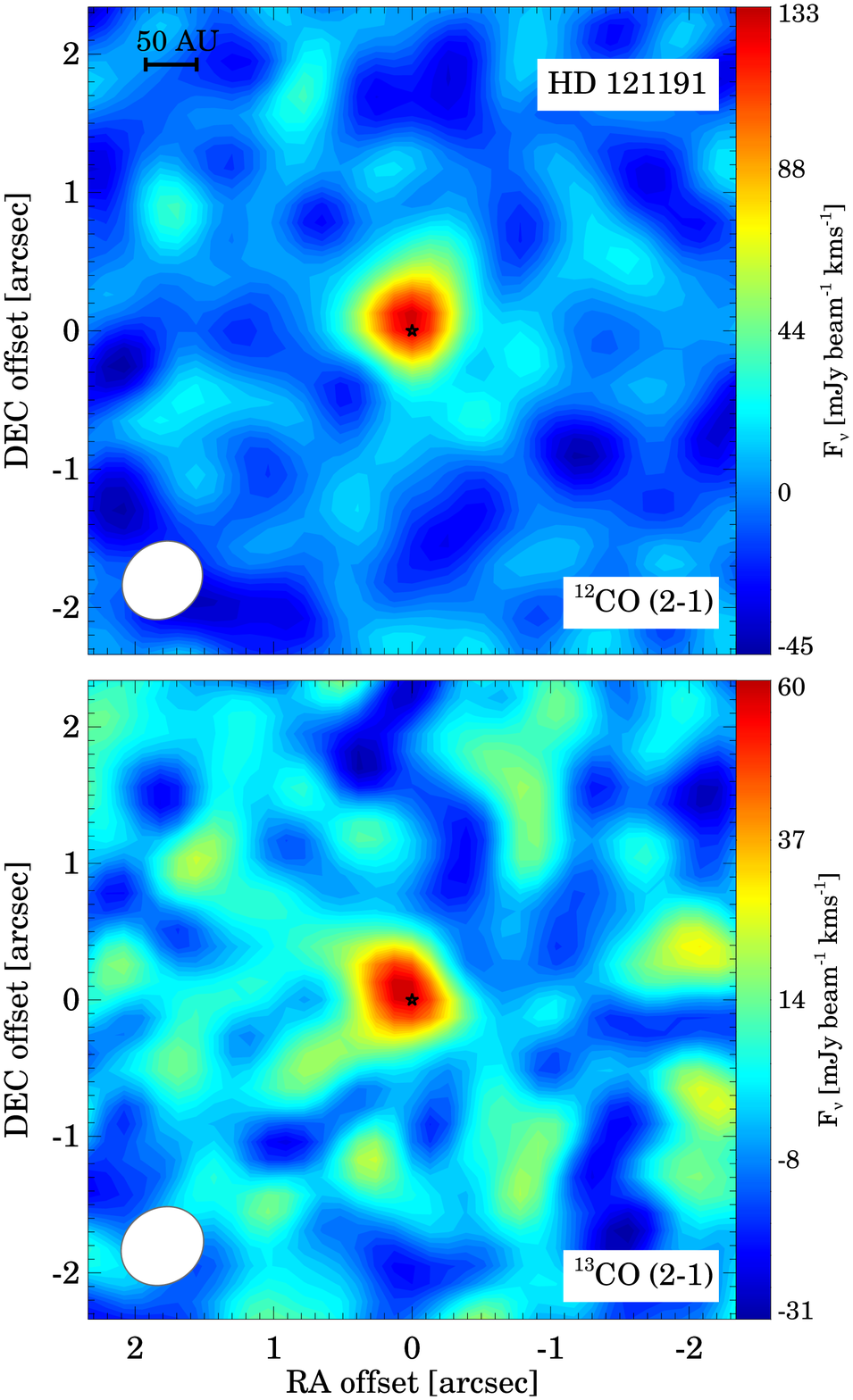}
\includegraphics[angle=0,scale=.3]{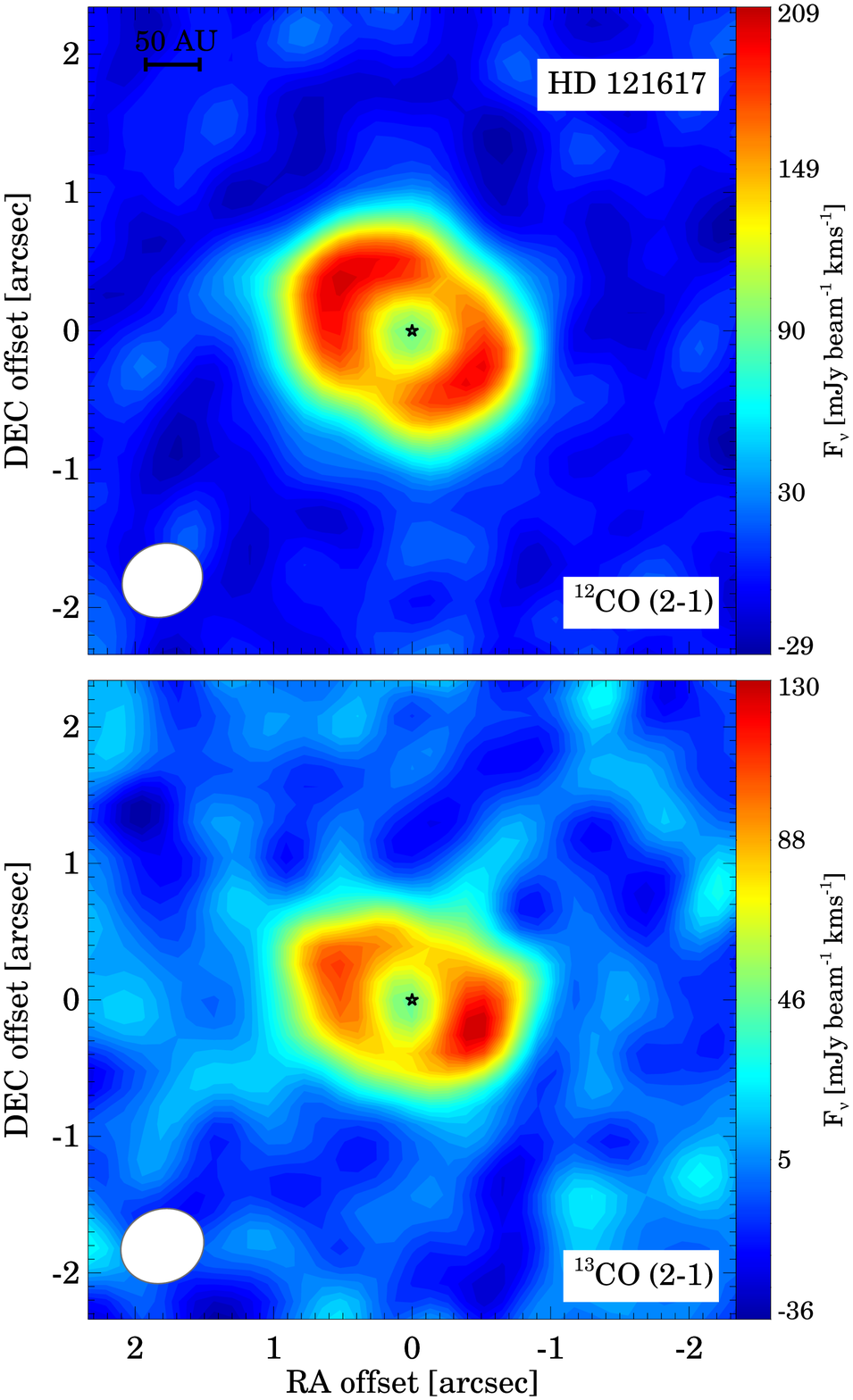}
\includegraphics[angle=0,scale=.3]{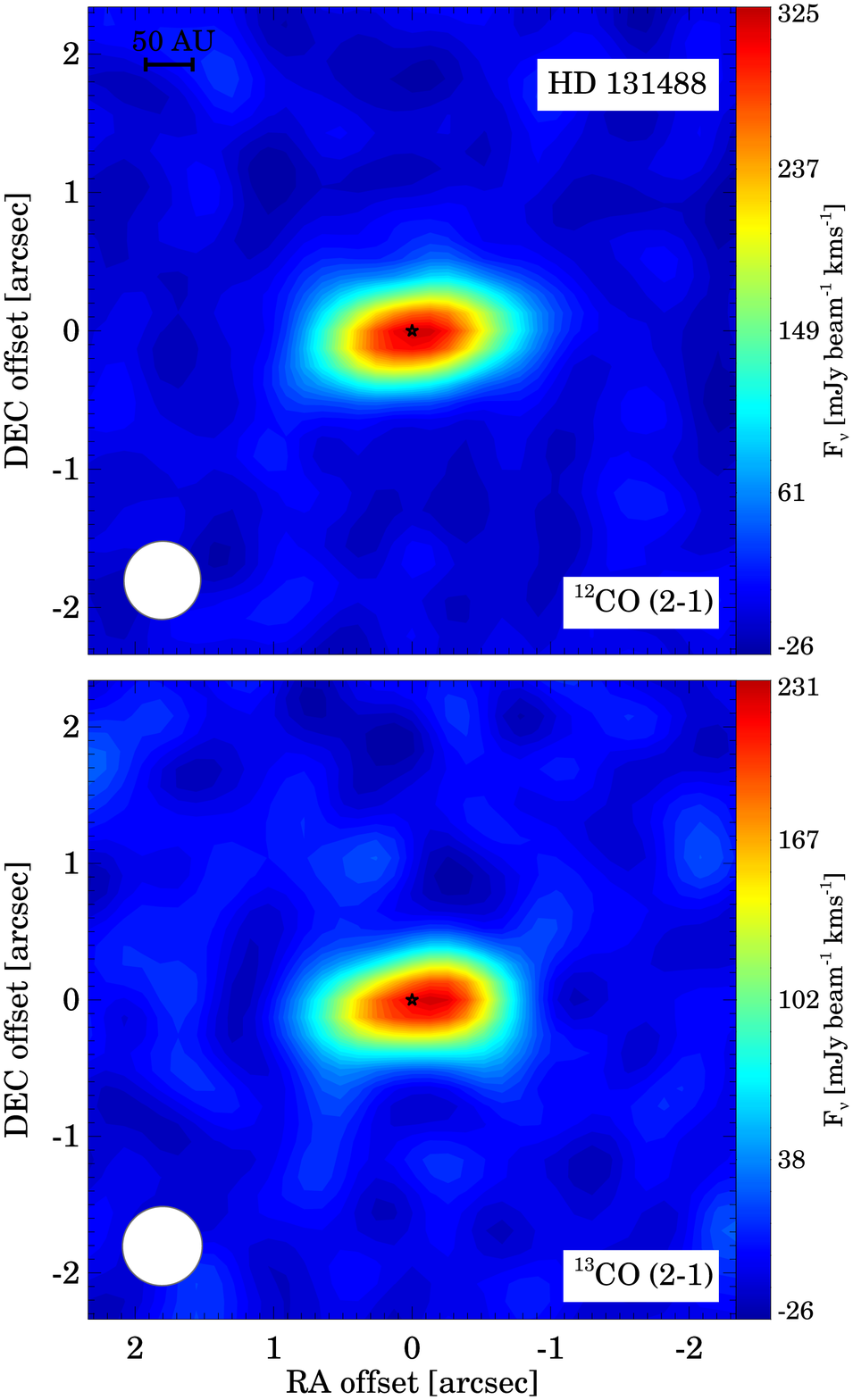}
\hspace*{-11mm}
\includegraphics[angle=0,scale=.31]{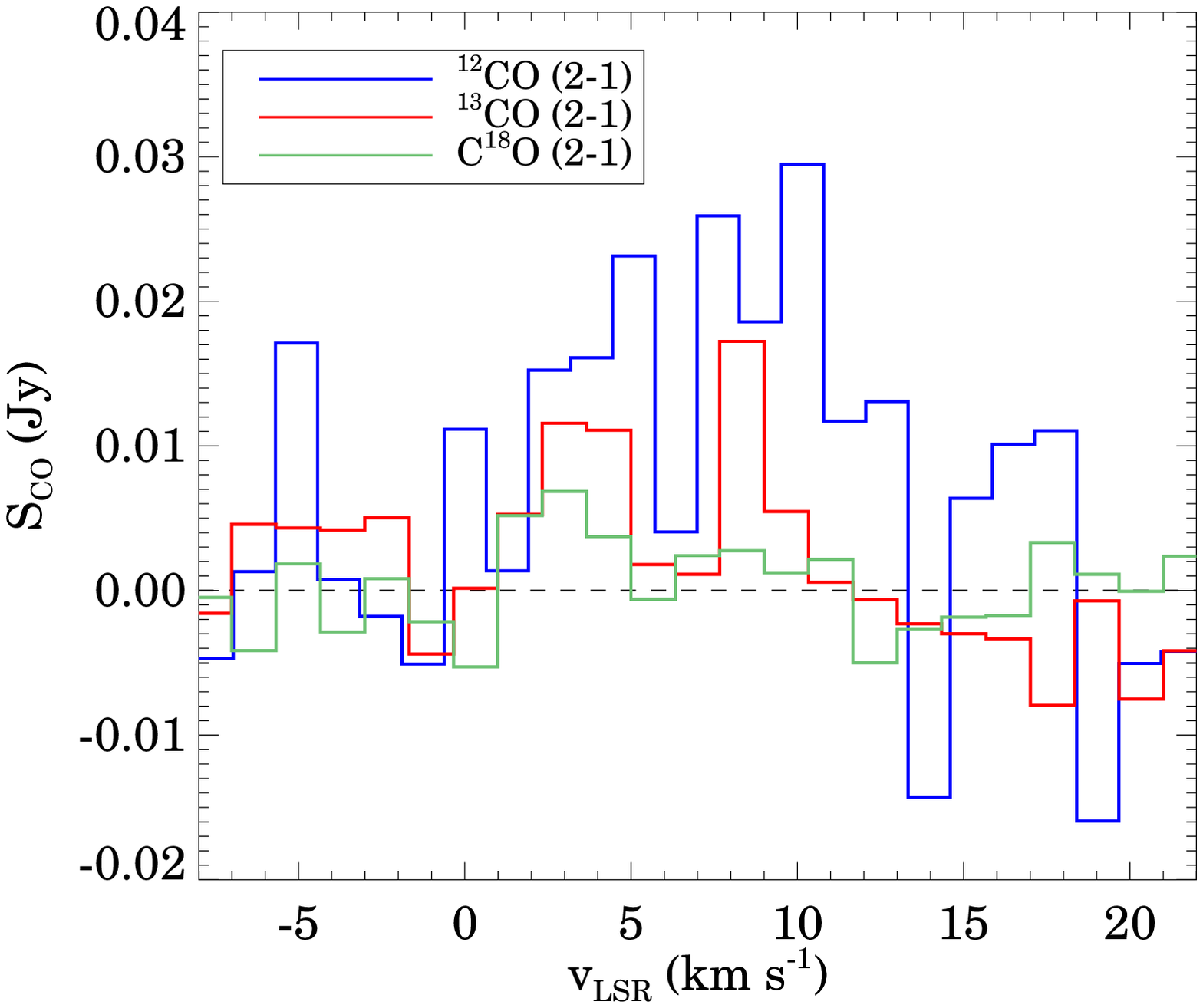}
\hspace*{-2mm}
\includegraphics[angle=0,scale=.31]{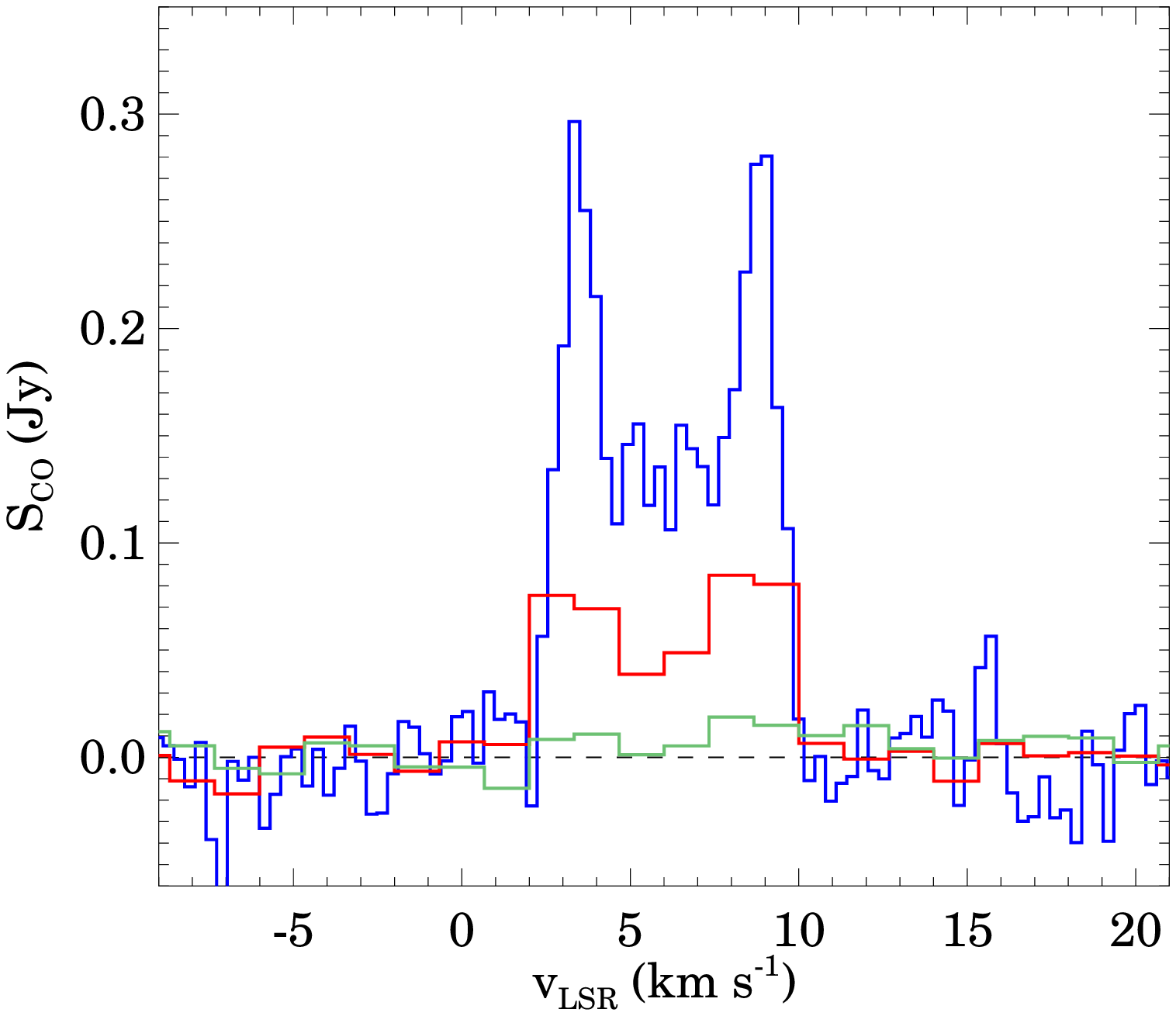}
\hspace*{-2mm}
\includegraphics[angle=0,scale=.31]{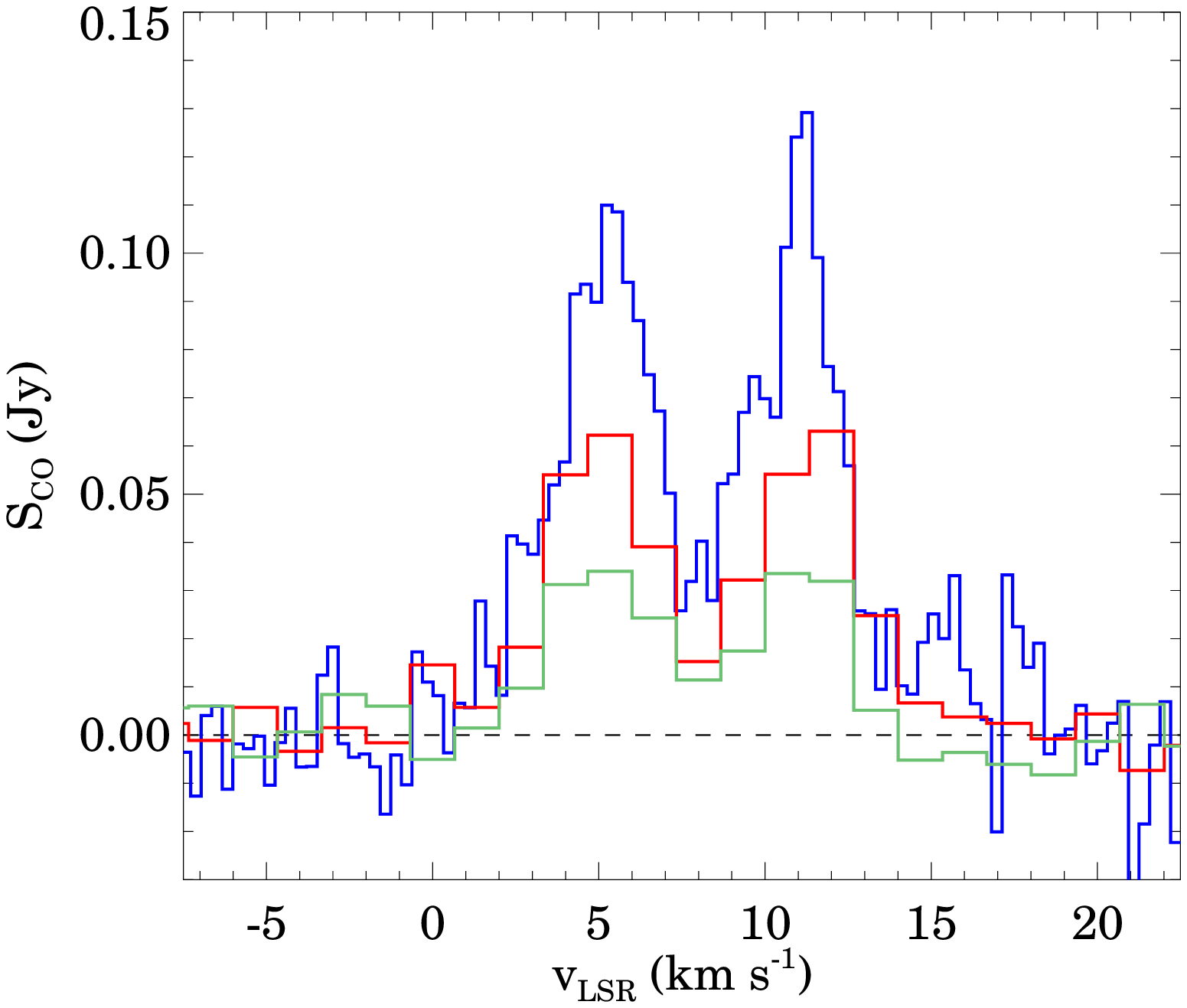}
\end{center}
\caption{ALMA line data for the three CO-bearing disks. The first
  and second rows show the $^{12}$CO and $^{13}$CO zeroth moment
  maps, respectively, while the third row displays the CO spectra.
\label{fig:line}}
\end{figure*}

\paragraph{Line data.} 
We detected
significant CO emission in HD~121191, HD~121617, and HD~131488. We
detected all three CO isotopologues in all three targets, except for
the C$^{18}$O in HD~121191. Integrating for the velocity ranges where
significant line emission can be seen, we calculated zeroth moment
maps (Fig.~\ref{fig:line} shows these for $^{12}$CO and
$^{13}$CO). 
Next, we determined an aperture which contains all CO
flux, measured the integrated line fluxes, and 
calculated the spectra (Fig.~\ref{fig:line}). The spectra
with higher signal-to-noise ratio show clear Keplerian profiles. 
For the
non-detections, we used a velocity range of $\pm$5\,km\,s$^{-1}$ of
the stellar radial velocity and an aperture radius corresponding to 100\,au to estimate 3$\sigma$ CO flux upper limits. Our results are
displayed in Tab.~\ref{tab:log}.

For the three detected sources, we estimated the {disk-averaged} optical depths of the
different CO isotopologues, from the measured line ratios. {Assuming  identical excitation temperatures, {emitting regions, and line widths}
for all isotopologues the {ratio of the $^{12}$CO and C$^{18}$O 
integrated brightness temperatures} 
can be approximated as 
$ \frac{1-e^{-\tau_{12}}}{1-e^{-\tau_{18}}}$
and similar formulae can be used for the $^{13}$CO to C$^{18}$O or $^{12}$CO to C$^{13}$O line ratios 
\citep[see, e.g.,][]{lyo2011}.}Adopting isotope ratios typical for the local interstellar matter,
i.e. [$^{12}$C]/[$^{13}$C]=77 and [$^{16}$O]/[$^{18}$O]=560
\citep{wilson1994}, we found the $^{12}$CO line to be highly optically
thick ($\tau_{12}\,{>}\,{30}$) in all three disks. For HD\,131488, even
the $^{13}$CO line is optically thick {with an optical depth of $\sim$5.7, 
while for HD\,121191 and  HD\,121617 this line was found to be optically thin.
Isotope selective photodissociation, and gas phase ion--molecule reactions
may lead to different isotopologue ratios in circumstellar disks than in the 
local interstellar matter \citep{visser2009,miotello2014}. 
For instance, due to these processes, the C$^{18}$O molecules are expected to be underabundant
by up to a factor of 10 \citep{miotello2014}. It may increase the calculated $^{12}$CO 
and $^{13}$CO optical depth, potentially questioning our conclusion of the optically thin 
$^{13}$CO emission in the case of HD\,121617.}

We used the peak flux values of the optically thick $^{12}$CO data
to estimate the gas brightness temperature, and obtained 6\,K,
10\,K, and 18\,K for HD~121191, HD~131488, and HD~121617,
respectively. Since the emitting region might be smaller than our beam
leading to beam dilution, these values should be considered as lower
limits. 

{Using an optically thin line of the isotopologue $^{y}$C$^{z}$O, the total 
$^{12}$CO mass can be obtained as: 
\begin{equation} 
M_{CO} = {4 \pi m d^2} \frac{S_{ul}}{x_u h \nu_{ul} A_{ul}} f_{^{12}C^{16}O/^{y}C^{z}O},    
\end{equation}
where $m$ is the mass of the {$^{12}$CO} molecule, $d$ is the distance of the object, $h$
is the Planck constant. $\nu_{ul}$ and $A_{ul}$ are the rest frequency and the
Einstein coefficient  for the given transition between the $u$ upper and $l$
lower levels, $S_{ul}$ is the observed integrated line flux,  $x_u$ is the
fractional population of the upper level{; all these parameters are related 
to the specific $^{y}C^{z}O$ isotopologue.} $f_{\rm ^{12}C^{16}O/^{y}C^{z}O}$ is
the abundance  of the $^{12}$CO molecule relative to the specific CO
isotopologue (we adopted local interstellar ratios, see above). In our observations we
detected  the J=2--1 transition of the different CO isotopologues. By assuming
local thermodynamical equilibrium (LTE) and adopting 20\,K for the gas
temperature, we determined $x_2$  from the Boltzmann equation. We used C$^{18}$O
for HD~131488 and $^{13}$CO for HD~121191, yielding total CO masses of 
{8.9$\pm$1.5$\times$10$^{-2}$\,M$_\oplus$} and 
{2.7$\pm$0.9$\times$10$^{-3}$\,M$_\oplus$,} respectively. In
the case of HD\,121617 both the $^{13}$CO and C$^{18}$O lines may be optically
thin, providing concordant CO gas mass estimates of {1.8$\pm$0.3$\times$10$^{-2}$\,M$_\oplus$}
and {2.0$\pm$0.7$\times$10$^{-2}$\,M$_\oplus$,} respectively.}  

{The quoted uncertainties
account only for the errors in the measured line fluxes and in the distances.
There could be, however, other sources of uncertainty. 
Similarly to the dust mass estimate, the gas temperature was 
characterized by a single value instead of a more realistic temperature 
distribution.  
Assuming LTE, the
fractional population of the upper J=2 level has a maximum at around our adopted
gas temperature of 20\,K.  
Taking temperatures lower than $\sim$14\,K 
or higher than 20\,K we would get lower fractional populations, i.e. 
higher gas masses. Between gas temperatures of 8 and 48\,K, the 
mass estimate is at most 1.5$\times$ higher the one derived for 20\,K.}

{LTE requires a dense medium where the collisions of CO with other particles are
sufficiently frequent. In lower density environments, however, 
radiative excitation can dominate over collisional processes. Such a non-LTE 
situation can lead to weakly populated levels, even at the J=2 rotational 
level of the CO molecule, resulting in higher mass estimates.
Isotopologue abundance ratios different from the local interstellar values also affect
the results. The combination of all these effects may lead to somewhat 
higher values than our mass estimates in Tab.~\ref{tab:props}.} 

{Eight out of the other ten disks have been detected in CO lines.
For HD\,21997, $\beta$\,Pic, and HD\,131835, the mass estimates were taken from \citet{kospal2013}, 
\citet{matra2017}, and Mo\'or et al. in prep., respectively. 
For the other detected targets (49\,Cet, HD\,32297, HD\,110058, HD\,138813, HD\,156623) we used
their measured $^{12}$CO line fluxes (see Tab.~\ref{tab:props}), and by assuming LTE, we adopted
32\,K for 49\,Cet \citep{hughes2017}, and 20\,K for the other sources as the gas
excitation temperature in the CO mass calculation.
Taking into account that the $^{12}$CO line could be optically thick in these targets and 
considering the already mentioned caveats about CO mass estimates (see above) we considered our results 
as lower limits (Tab.~\ref{tab:props}).
}

\section{Discovery of three new gaseous debris disks} \label{discovery}

\paragraph{HD~131488} is an A1-type member of the $\sim$16\,Myr-old UCL
association. Based on its SED, the debris disk has a high fractional
luminosity and may contain two belts, the inner one being
exceptionally hot \citep{melis2013}. 
This system has the highest dust and {CO} gas mass among the known
gaseous debris disks. The dust emission is elongated
in the east-west direction, with a central depression. We fitted the
continuum visibilities with the \texttt{uvmultifit} task using a
Gaussian ring model \citep{marti-vidal2014}. The best-fit parameters
are: total flux of 2.9$\pm$0.1\,mJy, ring diameter of
1$\farcs$14$\pm$0$\farcs$04 (168$\pm$6\,au), inclination of
82{\degr}$\pm$3{\degr}, position angle of 96$\pm$1{\degr}, FWHM of the
ring thickness of 0$\farcs$30$\pm$0$\farcs$08 (44$\pm$11\,au). The
center of the model does not show significant offset from the stellar
position. 
The obtained ring may correspond to the cold{, outer} dust belt
inferred from the SED, {however, the measured ring radius is 2.7$\pm$0.3 times larger 
than the one computed by assuming blackbody grains based on the stellar luminosity 
and the dust temperature quoted in Tab.~\ref{tab:props}.  
Using a sample of 34 spatially 
resolved debris disks, \citet{pawellek2014} studied how the ratios of the measured disk 
radii to the blackbody radii ($\Gamma = R_{\rm meas}/R_{\rm BB}$) vary 
as a function of stellar luminosity. By placing our 
target in their figure 4b (where they calculated $\Gamma$ similarly to our analysis), 
we found that the obtained $\Gamma$ value
is higher than any of the other systems with similarly bright host stars, suggesting an 
overabundance of small grains that are hotter than blackbodies. 
}

\paragraph{HD~121617} is also an A1-type member of the $\sim$16\,Myr-old UCL
association. Its SED shows high fractional luminosity and is
consistent with a single ring model with $T_{\rm dust}$=105\,K
\citep{moor2011}. The continuum image shows an inclined ring-like
morphology. The \texttt{uvmultifit} procedure yielded the
following parameters: total flux of 1.9$\pm$0.2\,mJy, ring diameter of
1$\farcs$29$\pm$0$\farcs$12 (165$\pm$16\,au), inclination of
37{\degr}$\pm$13{\degr}, position angle of 43$\pm$19{\degr}, FWHM of
the ring thickness of 0$\farcs$44$\pm$0$\farcs$15 (57$\pm$19\,au).
{In this case, the ratio of the inferred radius to the blackbody radius 
is $\Gamma =$2.9$\pm$0.4. This $\Gamma$ value and a stellar luminosity that is higher than that 
of our previous target, makes HD\,121617 an even more prominent outlier in figure 4b
of \citet{pawellek2014}, also implying the presence of copious amount of small grains. 
Remarkably, by analyzing the characteristic grain sizes relative to the 
theoretical blowout size in 12 spatially resolved young debris disks, \citet{lieman-sifry2016} 
found that the CO-bearing systems in the sample had grain sizes on the small end of the
distribution, and two of them had grain sizes smaller than the blowout size.
}

\paragraph{HD~121191.} The position and space motion of this A5-type
star can be consistent with either the UCL ($\sim$16\,Myr) or LCC ($\sim$15\,Myr)
sub-groups of the Sco-Cen association \citep{melis2013}.
Based on its SED, it has a high fractional luminosity and it may also
have two belts, the inner one being very hot and bright
\citep{melis2013}. The weak 1.3\,mm continuum emission from this target is
unresolved, 
and {its peak} is offset from
the stellar position toward the SW ($\Delta$RA=$-$0\farcs16$\pm$0\farcs05; $\Delta$Dec=$-$0\farcs17$\pm$0\farcs05). 
This may be {an axisymmetric} ring whose
parts remain below the noise level, or may be a ring with highly
asymmetric brightness distribution, although contamination by a
background source cannot be excluded either. 
Both $^{12}$CO and $^{13}$CO were firmly detected as {compact}
sources centered on the stellar position. This disk has a lower dust
mass than the previous two targets.

\section{Molecular gas in bright debris disks}

\begin{figure*}
\begin{center}
\includegraphics[angle=0,scale=.5]{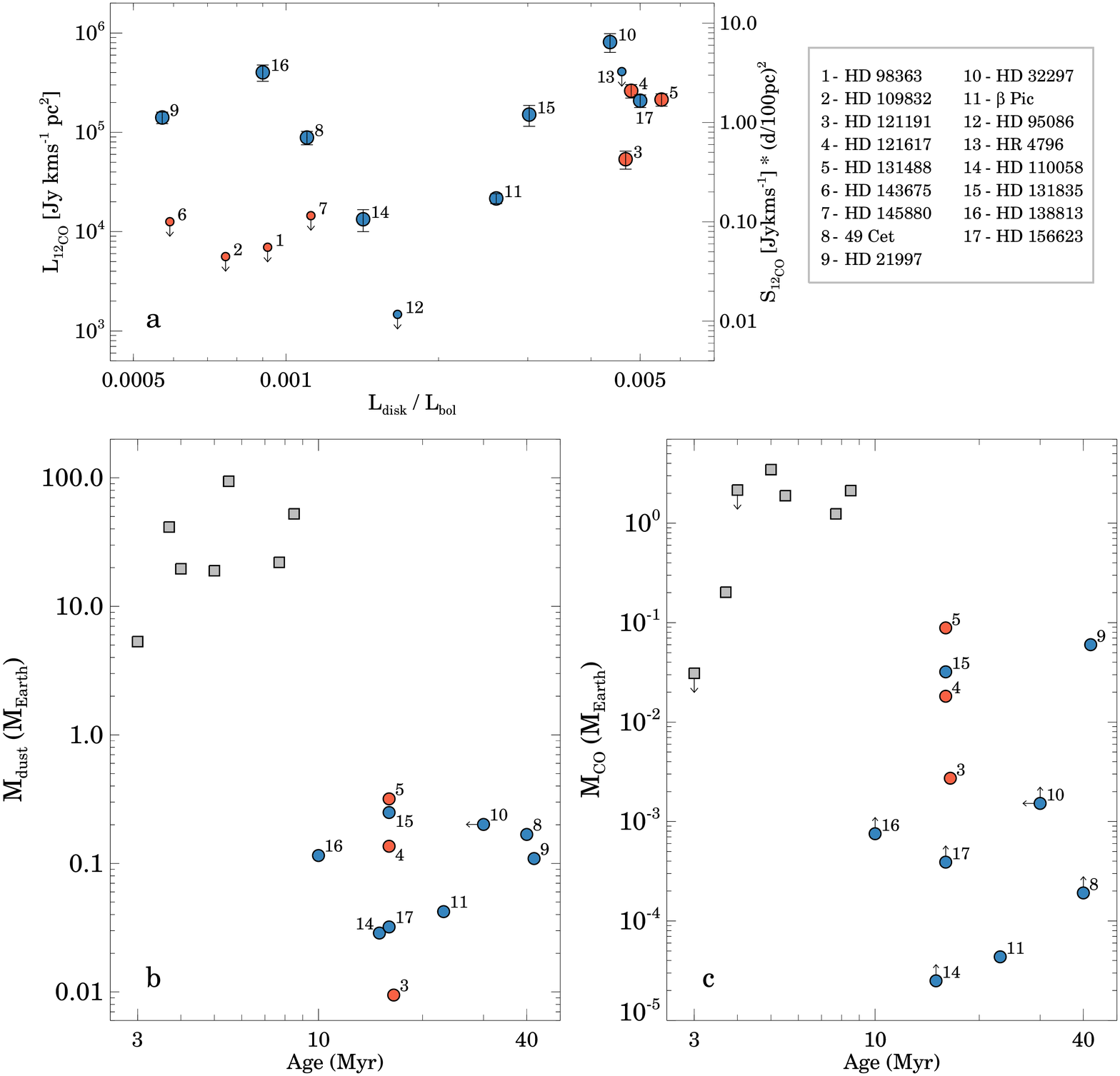} 
\end{center}
\caption{a) {$^{12}$CO 2--1 (3--2 for HR\,4796) line luminosity 
 for the selected 17 bright debris disks as a function of 
 fractional luminosity. For the sake of compatibility with 
 previous works, integrated line fluxes normalized to 100\,pc are 
 also drawn.} 
  Our current seven targets are
  marked by red symbols. Sources with no CO detection are displayed
  with smaller symbols. b) Dust masses of circumstellar disks derived
  from observed (sub)millimetre fluxes towards disks around some
  selected Herbig stars (gray squares), and CO-bearing debris disks.
  For consistency, millimeter fluxes of Herbig disks
  \citep[taken from][]{raman2006,isella2007,chapillon2008,oberg2010,ansdell2016,pericaud2017}
  were converted to dust masses by adopting an opacity of $\kappa_\nu =
  2.3\left(\frac{\nu}{{\rm 230\,GHz}}\right)^{0.7}$ and a dust
  temperature of 50\,K. c) CO gas masses derived from optically thin
  $^{12}$CO or CO isotopologue ($^{13}$CO, C$^{17}$O or C$^{18}$O) line
  observations for the same objects. To compute CO masses for Herbig
  sample, we assumed LTE and a gas excitation temperature of 50\,K.
  CO isotopologue data are from 
\citet{rosenfeld2013,guilloteau2013,guilloteau2016,ansdell2016,fedele2017}.\label{fig:disc}}
\end{figure*}

Until recently, only {$\beta$\,Pic and 49~Cet were known as CO-bearing debris
disks} \citep{vidal-madjar1994,zuckerman1995,roberge2000}. Later, more such objects were discovered, and the first
statistical studies could be done 
\citep{lieman-sifry2016, greaves2016,pericaud2017}. Our present list partly
incorporates these earlier samples, and extends them to a full list
of 17 {dust-rich} debris disks in the Solar neighborhood
(Sec.~\ref{sec:targetsel}). 
{In Fig.~\ref{fig:disc}a we display the $^{12}$CO 2--1
(or $^{12}$CO 3--2 for HR\,4796) line luminosities (or upper
limits), as a function of fractional luminosity for our sample. 
Line luminosities of detected CO-bearing disks span almost two orders of magnitude, the 
brightest disks have luminosities comparable to those of fainter Herbig Ae and T\,Tauri disks
\citep{ansdell2016,pericaud2017}. Since the sensitivity of the HR\,4796 observation is nearly two orders of magnitude {worse than that} of the other 
measurements, we discard this object from the following {analysis,} reducing the size of our statistical sample to 16.}
{For the other objects, adopting the highest upper limit (source no. 7) 
in Fig.~\ref{fig:disc}a, we derived a detectability threshold of $\sim$1.4$\times$10$^{4}$ Jy~km\,s$^{-1}$pc$^2$ for the $^{12}$CO 2--1 line luminosity.}
With our three new discoveries, we found 11
disks in this sample that harbor CO gas, resulting in a very high
detection rate of {68.8$^{+8.9}_ {-13.1}$\%.} Because of the small sample, 
we computed the uncertainties {(corresponding to 68\% confidence interval)} 
using the binomial distribution approach {proposed by} \citet{burgasser2003}. 
{Our result} indicates that the presence of CO gas in {dust-rich} debris disks 
around young A-type stars is likely more the rule than the exception.
{The obtained incidence rate of 11/16 is valid above our detectability 
threshold for the $^{12}$CO J=2--1 line luminosities. 
Nevertheless we cannot rule out that all of our targets harbor CO gas at some level.
Remarkably, as Fig.~\ref{fig:disc}a shows, apart from HR\,4796, all disks with $L_{\rm disk}/L_{\rm bol} >
2\times10^{-3}$,  
contain detectable levels of CO gas.}

Dividing our list into two
subsamples that contain stars having lower and higher
luminosity than 10\,L$_\odot$, we get CO detection rates of 
62.5$^{+12.8}_{-17.9}$\% (5/8) and {75.0$^{+9.1}_{-19.3}$\% (6/8)},
respectively. This implies that there is {no evidence in support of a}
correlation with stellar luminosity within the A-type sample.

To compare with debris disks around later type stars, we 
collected previous ALMA results for those young (10--50\,Myr) 
debris disks that encircle F- and G-type stars, but otherwise 
fulfill our original selection criteria (Sec.~\ref{sec:targetsel}) {and their ALMA measurements 
achieved the same detection threshold 
($\sim$1.4$\times$10$^{4}$ Jy~km\,s$^{-1}$pc$^2$) as that of our A-type sample.}
We found 16 such systems: 14 belonging
to the Sco-Cen association \citep{lieman-sifry2016}, and two young
moving group members, HD\,61005 and HD\,181327
\citep{olofsson2016,marino2016}. 
From the FG stars, two objects {were found to} harbor
CO gas (HD~181327 and HD~146897). 
{However, with its $^{12}$CO 2--1 line luminosity of 
$\sim$10$^3$\,Jy~km~s$^{-1}$pc$^2$ \citep[based on][]{marino2016}, 
HD~181327 is at least an order of magnitude fainter than any of the other known gaseous disks from the  
studied samples. Its detection was only possible because of its proximity and the long exposure time.} 
{Such a faint CO disk would not be detectable at any objects in the A-type sample, 
therefore this source was discarded from the following analysis, yielding a detection rate 
of 6.7$^{+12.5}_{-2.2}$\% for disks around young FG-type stars.}  
By applying a Fisher's exact
test to compare the occurrence of CO-bearing disks 
in the two samples, we obtained a $p$ value of {6.4$\times$10$^{-4}$,} implying a statistically significant
difference between {dust-rich} debris disks around young A-type and FG-type stars 
{above our detectability threshold.} 
{This suggests that either the incidence of CO gas is truly lower around young FG-type star than in
A-type systems, or the CO line luminosities are systematically lower. The reason for the latter may be
related to lower CO gas content and/or weaker population of the J=2 rotational level due to less
efficient excitation (in optically {thin} case), or systematically lower CO temperatures and/or disk
sizes (in optically {thick} case).}

In Fig.~\ref{fig:disc}bc, we plotted $M_{\rm dust}$ and $M_{\rm CO}$
versus age for CO-bearing debris disks. To study the transition between
primordial and debris disks we also plotted some Herbig Ae/Fe systems 
with CO isotopologue observations from the literature. There is a
pronounced drop in dust mass from primordial to debris disks
\citep[see also,][]{roccatagliata2009,panic2013}. In {CO} gas mass, however, debris disks
show a significantly larger spread. Four gaseous debris
disks -- HD\,121617 and HD\,131488  from our current sample, as well as
HD\,21997 and HD\,131835 -- harbor comparable amounts of CO to that of {the
two least massive Herbig disks in the figure.} For these disks, our ALMA CO isotopologue 
observations revealed 30--250$\times$ more CO gas than that calculated 
from $^{12}$CO measurements (using an optically thin assumption),
highlighting the importance of less abundant molecules in reliable mass
estimates. 

While the gas in the disk of $\beta$\,Pic, and around the
two F-type hosts is secondary \citep{kral2017,marino2016,matra2017}, 
the presence of significantly larger amounts of CO gas at HD\,21997
and HD\,131835 indicates that these systems may have hybrid disks,
where the gas is {leftover from the primordial phase} (\citealt{kospal2013,kral2017},
Mo\'or et al.~in prep.). The {likely} existence of hybrid disks {would be} a strong
indication that in some systems the evolution of dust and gas, at least
in the outer disk, is decoupled. By exhibiting CO masses similar to that
of HD\,21997 and HD\,131835, the three newly discovered CO-bearing
disks, particularly HD\,121617 and HD\,131488, could be prime hybrid disk
candidates. The origin of gas
in these systems, along with their detailed morphological analysis, will
be the topic of a next paper. Four gaseous systems, 49\,Cet, HD\,32297,
HD\,138813, and HD\,156623 have similar $^{12}$CO 2--1 
{line luminosity to the above subsample,}
therefore their current CO mass estimates should be considered as lower limits. 
The origin of gas in these systems is not known.
Further observations at CO isotopologue lines might reveal that some of them may 
exhibit hybrid nature as well.
{If the disks with the highest $^{12}$CO line luminosities in Fig.~\ref{fig:disc}a 
indeed turn out to be hybrid disks, it would indicate a strong link between the hybrid disk 
phenomenon and young A-type stars.
If this link is exclusive, then it could help to understand why we detect a lower number of 
gaseous debris disks around FG-type stars above our current detectability threshold.}


\acknowledgments

We acknowledge the referee for insightful comments that
have improved the paper. We appreciate Ivan Mart\'\i -Vidal for 
helpful advices on the \texttt{uvmultifit} tool.
This paper makes use of the following ALMA data:
ADS/JAO.ALMA\#2015.1.01243.S. ALMA is a partnership of ESO
(representing its member states), NSF (USA) and NINS (Japan), together
with NRC (Canada) and NSC and ASIAA (Taiwan) and KASI (Republic of
Korea), in cooperation with the Republic of Chile. The Joint ALMA
Observatory is operated by ESO, AUI/NRAO and NAOJ. This work was
supported by the Momentum grant of the MTA CSFK Lend\"ulet Disk
Research Group and Hungarian OTKA grant K119993. A.M.H. gratefully
acknowledges support from NSF grant AST-1412647. Este trabajo
conta con el apoyo de CONICYT Programa de Astronomia Fondo
ALMA-CONICYT, cargo proyectos numeros: 31140024 y 31AS002.

\vspace{5mm}
\facilities{ALMA}

\bibliographystyle{yahapj}

\begin{thebibliography}{}
\expandafter\ifx\csname natexlab\endcsname\relax\def\natexlab#1{#1}\fi
\providecommand{\url}[1]{\href{#1}{#1}}
\bibitem[Alexander et al.(2014)]{alexander2014} Alexander, R., Pascucci, I., Andrews, S., Armitage, P., \& Cieza, L.\ 2014, Protostars and Planets VI, ed. H. Beuther et al. (Tucson, AZ: Univ. Arizona Press), 475 

\bibitem[{{Andrews} {et~al.}(2013){Andrews}, {Rosenfeld}, {Kraus}, \&
  {Wilner}}]{andrews2013}
{Andrews}, S.~M., {Rosenfeld}, K.~A., {Kraus}, A.~L., \& {Wilner}, D.~J. 2013,
  \apj, 771, 129

\bibitem[{{Ansdell} {et~al.}(2016){Ansdell}, {Williams}, {van der Marel},
  {Carpenter}, {Guidi}, {Hogerheijde}, {Mathews}, {Manara}, {Miotello},
  {Natta}, {Oliveira}, {Tazzari}, {Testi}, {van Dishoeck}, \& {van
  Terwisga}}]{ansdell2016}
{Ansdell}, M., {Williams}, J.~P., {van der Marel}, N., {et~al.} 2016, \apj,
  828, 46

\bibitem[{{Ballering} {et~al.}(2013){Ballering}, {Rieke}, {Su}, \&
  {Montiel}}]{ballering2013}
{Ballering}, N.~P., {Rieke}, G.~H., {Su}, K.~Y.~L., \& {Montiel}, E. 2013,
  \apj, 775, 55

\bibitem[Barrado y Navascu{\'e}s et al.(1999)]{barrado1999} 
Barrado y Navascu{\'e}s, D., Stauffer, J.~R., Song, I., \& Caillault, J.-P.\ 1999, \apjl, 520, L123 

\bibitem[Bell et al.(2015)]{bell2015} Bell, C.~P.~M., Mamajek, E.~E., \& Naylor, T.\ 2015, 
\mnras, 454, 593

\bibitem[{{Burgasser} {et~al.}(2003){Burgasser}, {Kirkpatrick}, {Reid},
  {Brown}, {Miskey}, \& {Gizis}}]{burgasser2003}
{Burgasser}, A.~J., {Kirkpatrick}, J.~D., {Reid}, I.~N., {et~al.} 2003, \apj,
  586, 512

\bibitem[{{Chapillon} {et~al.}(2008){Chapillon}, {Guilloteau}, {Dutrey}, \&
  {Pi{\'e}tu}}]{chapillon2008}
{Chapillon}, E., {Guilloteau}, S., {Dutrey}, A., \& {Pi{\'e}tu}, V. 2008, \aap,
  488, 565

\bibitem[{{Chen} {et~al.}(2014){Chen}, {Mittal}, {Kuchner}, {Forrest}, {Lisse},
  {Manoj}, {Sargent}, \& {Watson}}]{chen2014}
{Chen}, C.~H., {Mittal}, T., {Kuchner}, M., {et~al.} 2014, \apjs, 211, 25

\bibitem[{{Dent} {et~al.}(2014){Dent}, {Wyatt}, {Roberge}, {Augereau},
  {Casassus}, {Corder}, {Greaves}, {de Gregorio-Monsalvo}, {Hales}, {Jackson},
  {Hughes}, {Lagrange}, {Matthews}, \& {Wilner}}]{dent2014}
{Dent}, W.~R.~F., {Wyatt}, M.~C., {Roberge}, A., {et~al.} 2014, Science, 343,
  1490

\bibitem[{{Fedele} {et~al.}(2017){Fedele}, {Carney}, {Hogerheijde}, {Walsh},
  {Miotello}, {Klaassen}, {Bruderer}, {Henning}, \& {van
  Dishoeck}}]{fedele2017}
{Fedele}, D., {Carney}, M., {Hogerheijde}, M.~R., {et~al.} 2017, \aap, 600, A72

\bibitem[{{Greaves} {et~al.}(2016){Greaves}, {Holland}, {Matthews}, {Marshall},
  {Dent}, {Woitke}, {Wyatt}, {Matr{\`a}}, \& {Jackson}}]{greaves2016}
{Greaves}, J.~S., {Holland}, W.~S., {Matthews}, B.~C., {et~al.} 2016, \mnras,
  461, 3910

\bibitem[{{Guilloteau} {et~al.}(2013){Guilloteau}, {Di Folco}, {Dutrey},
  {Simon}, {Grosso}, \& {Pi{\'e}tu}}]{guilloteau2013}
{Guilloteau}, S., {Di Folco}, E., {Dutrey}, A., {et~al.} 2013, \aap, 549, A92

\bibitem[{{Guilloteau} {et~al.}(2016){Guilloteau}, {Reboussin}, {Dutrey},
  {Chapillon}, {Wakelam}, {Pi{\'e}tu}, {Di Folco}, {Semenov}, \&
  {Henning}}]{guilloteau2016}
{Guilloteau}, S., {Reboussin}, L., {Dutrey}, A., {et~al.} 2016, \aap, 592, A124

\bibitem[{{Hales} {et~al.}(2014){Hales}, {De Gregorio-Monsalvo}, {Montesinos},
  {Casassus}, {Dent}, {Dougados}, {Eiroa}, {Hughes}, {Garay}, {Mardones},
  {M{\'e}nard}, {Palau}, {P{\'e}rez}, {Phillips}, {Torrelles}, \&
  {Wilner}}]{hales2014}
{Hales}, A.~S., {De Gregorio-Monsalvo}, I., {Montesinos}, B., {et~al.} 2014,
  \aj, 148, 47


\bibitem[Hoogerwerf(2000)]{hoogerwerf2000} Hoogerwerf, R.\ 2000, \mnras, 313, 43 

\bibitem[{{Hughes} {et~al.}(2008){Hughes}, {Wilner}, {Kamp}, \&
  {Hogerheijde}}]{hughes2008}
{Hughes}, A.~M., {Wilner}, D.~J., {Kamp}, I., \& {Hogerheijde}, M.~R. 2008,
  \apj, 681, 626

\bibitem[{{Hughes} {et~al.}(2017){Hughes}, {Lieman-Sifry}, {Flaherty}, {Daley},
  {Roberge}, {K{\'o}sp{\'a}l}, {Mo{\'o}r}, {Kamp}, {Wilner}, {Andrews},
  {Kastner}, \& {{\'A}brah{\'a}m}}]{hughes2017}
{Hughes}, A.~M., {Lieman-Sifry}, J., {Flaherty}, K.~M., {et~al.} 2017, \apj,
  839, 86

\bibitem[{{Isella} {et~al.}(2007){Isella}, {Testi}, {Natta}, {Neri}, {Wilner},
  \& {Qi}}]{isella2007}
{Isella}, A., {Testi}, L., {Natta}, A., {et~al.} 2007, \aap, 469, 213

\bibitem[Kalas(2005)]{kalas2005} Kalas, P.\ 2005, \apjl, 635, L169 

\bibitem[{{K{\'o}sp{\'a}l} {et~al.}(2013){K{\'o}sp{\'a}l}, {Mo{\'o}r},
  {Juh{\'a}sz}, {{\'A}brah{\'a}m}, {Apai}, {Csengeri}, {Grady}, {Henning},
  {Hughes}, {Kiss}, {Pascucci}, \& {Schmalzl}}]{kospal2013}
{K{\'o}sp{\'a}l}, {\'A}., {Mo{\'o}r}, A., {Juh{\'a}sz}, A., {et~al.} 2013,
  \apj, 776, 77

\bibitem[Kral et al.(2017)]{kral2017} Kral, Q., Matr{\`a}, L., Wyatt, M.~C., \& Kennedy, G.~M.\ 2017, \mnras, 469, 521

\bibitem[{{Lieman-Sifry} {et~al.}(2016){Lieman-Sifry}, {Hughes}, {Carpenter},
  {Gorti}, {Hales}, \& {Flaherty}}]{lieman-sifry2016}
{Lieman-Sifry}, J., {Hughes}, A.~M., {Carpenter}, J.~M., {et~al.} 2016, \apj,
  828, 25

\bibitem[{{Lindegren} {et~al.}(2016){Lindegren}, {Lammers}, {Bastian},
  {Hern{\'a}ndez}, {Klioner}, {Hobbs}, {Bombrun}, {Michalik}, {Ramos-Lerate},
  {Butkevich}, {Comoretto}, {Joliet}, {Holl}, {Hutton}, {Parsons},
  {Steidelm{\"u}ller}, {Abbas}, {Altmann}, {Andrei}, {Anton}, {Bach},
  {Barache}, {Becciani}, {Berthier}, {Bianchi}, {Biermann}, {Bouquillon},
  {Bourda}, {Br{\"u}semeister}, {Bucciarelli}, {Busonero}, {Carlucci},
  {Casta{\~n}eda}, {Charlot}, {Clotet}, {Crosta}, {Davidson}, {de Felice},
  {Drimmel}, {Fabricius}, {Fienga}, {Figueras}, {Fraile}, {Gai}, {Garralda},
  {Geyer}, {Gonz{\'a}lez-Vidal}, {Guerra}, {Hambly}, {Hauser}, {Jordan},
  {Lattanzi}, {Lenhardt}, {Liao}, {L{\"o}ffler}, {McMillan}, {Mignard}, {Mora},
  {Morbidelli}, {Portell}, {Riva}, {Sarasso}, {Serraller}, {Siddiqui}, {Smart},
  {Spagna}, {Stampa}, {Steele}, {Taris}, {Torra}, {van Reeven}, {Vecchiato},
  {Zschocke}, {de Bruijne}, {Gracia}, {Raison}, {Lister}, {Marchant},
  {Messineo}, {Soffel}, {Osorio}, {de Torres}, \& {O'Mullane}}]{tgas}
{Lindegren}, L., {Lammers}, U., {Bastian}, U., {et~al.} 2016, \aap, 595, A4

\bibitem[Lyo et al.(2011)]{lyo2011} Lyo, A.-R., Ohashi, N., Qi, C., Wilner, D.~J., \& Su, Y.-N.\ 2011, \aj, 142, 151 

\bibitem[Mamajek \& Bell(2014)]{mamajek2014} Mamajek, E.~E., \& Bell, C.~P.~M.\ 2014, \mnras, 445, 2169 

\bibitem[{{Marino} {et~al.}(2016){Marino}, {Matr{\`a}}, {Stark}, {Wyatt},
  {Casassus}, {Kennedy}, {Rodriguez}, {Zuckerman}, {Perez}, {Dent}, {Kuchner},
  {Hughes}, {Schneider}, {Steele}, {Roberge}, {Donaldson}, \&
  {Nesvold}}]{marino2016}
{Marino}, S., {Matr{\`a}}, L., {Stark}, C., {et~al.} 2016, \mnras, 460, 2933

\bibitem[Marino et al.(2017)]{marino2017} Marino, S., Wyatt, M.~C., Pani{\'c}, O., 
et al.\ 2017, \mnras, 465, 2595 

\bibitem[{{Mart{\'{\i}}-Vidal} {et~al.}(2014){Mart{\'{\i}}-Vidal}, {Vlemmings},
  {Muller}, \& {Casey}}]{marti-vidal2014}
{Mart{\'{\i}}-Vidal}, I., {Vlemmings}, W.~H.~T., {Muller}, S., \& {Casey}, S.
  2014, \aap, 563, A136

\bibitem[{{Matr{\`a}} {et~al.}(2017a){Matr{\`a}}, {Dent}, {Wyatt}, {Kral},
  {Wilner}, {Pani{\'c}}, {Hughes}, {de Gregorio-Monsalvo}, {Hales}, {Augereau},
  {Greaves}, \& {Roberge}}]{matra2017}
{Matr{\`a}}, L., {Dent}, W.~R.~F., {Wyatt}, M.~C., {et~al.} 2017, \mnras, 464,
  1415

\bibitem[Matr{\`a} et al.(2017b)]{matra2017b} Matr{\`a}, L., MacGregor, M.~A., Kalas, P., et al.\ 2017, arXiv:1705.05868

\bibitem[{{Matthews} {et~al.}(2014){Matthews}, {Krivov}, {Wyatt}, {Bryden}, \&
  {Eiroa}}]{matthews2014}
{Matthews}, B.~C., {Krivov}, A.~V., {Wyatt}, M.~C., {Bryden}, G., \& {Eiroa},
  C. 2014, Protostars and Planets VI, 521

\bibitem[{{McMullin} {et~al.}(2007){McMullin}, {Waters}, {Schiebel}, {Young},
  \& {Golap}}]{mcmullin2007}
{McMullin}, J.~P., {Waters}, B., {Schiebel}, D., {Young}, W., \& {Golap}, K.
  2007, in Astronomical Society of the Pacific Conference Series, Vol. 376,
  Astronomical Data Analysis Software and Systems XVI, ed. R.~A. {Shaw},
  F.~{Hill}, \& D.~J. {Bell}, 127

\bibitem[{{Meeus} {et~al.}(2012){Meeus}, {Montesinos}, {Mendigut{\'{\i}}a},
  {Kamp}, {Thi}, {Eiroa}, {Grady}, {Mathews}, {Sandell}, {Martin-Za{\"i}di},
  {Brittain}, {Dent}, {Howard}, {M{\'e}nard}, {Pinte}, {Roberge},
  {Vandenbussche}, \& {Williams}}]{meeus2012}
{Meeus}, G., {Montesinos}, B., {Mendigut{\'{\i}}a}, I., {et~al.} 2012, \aap,
  544, A78

\bibitem[{{Melis} {et~al.}(2013){Melis}, {Zuckerman}, {Rhee}, {Song}, {Murphy},
  \& {Bessell}}]{melis2013}
{Melis}, C., {Zuckerman}, B., {Rhee}, J.~H., {et~al.} 2013, \apj, 778, 12

\bibitem[Miotello et al.(2014)]{miotello2014} Miotello, A., Bruderer, S., \& van Dishoeck, E.~F.\ 2014, \aap, 572, A96 

\bibitem[Miyake \& Nakagawa(1993)]{miyake1993} Miyake, K., \& Nakagawa, Y.\ 1993, \icarus, 106, 20

\bibitem[Mo{\'o}r et al.(2006)]{moor2006} Mo{\'o}r, A., {\'A}brah{\'a}m, P., 
Derekas, A., et al.\ 2006, \apj, 644, 525 

\bibitem[{{Mo{\'o}r} {et~al.}(2011){Mo{\'o}r}, {{\'A}brah{\'a}m}, {Juh{\'a}sz},
  {Kiss}, {Pascucci}, {K{\'o}sp{\'a}l}, {Apai}, {Henning}, {Csengeri}, \&
  {Grady}}]{moor2011}
{Mo{\'o}r}, A., {{\'A}brah{\'a}m}, P., {Juh{\'a}sz}, A., {et~al.} 2011, \apjl,
  740, L7

\bibitem[{{Mo{\'o}r} {et~al.}(2015{\natexlab{a}}){Mo{\'o}r}, {Henning},
  {Juh{\'a}sz}, {{\'A}brah{\'a}m}, {Balog}, {K{\'o}sp{\'a}l}, {Pascucci},
  {Szab{\'o}}, {Vavrek}, {Cur{\'e}}, {Csengeri}, {Grady}, {G{\"u}sten}, \&
  {Kiss}}]{moor2015b}
{Mo{\'o}r}, A., {Henning}, T., {Juh{\'a}sz}, A., {et~al.} 2015{\natexlab{a}},
  \apj, 814, 42

\bibitem[{{Mo{\'o}r} {et~al.}(2015{\natexlab{b}}){Mo{\'o}r}, {K{\'o}sp{\'a}l},
  {{\'A}brah{\'a}m}, {Apai}, {Balog}, {Grady}, {Henning}, {Juh{\'a}sz}, {Kiss},
  {Krivov}, {Pawellek}, \& {Szab{\'o}}}]{moor2015a}
{Mo{\'o}r}, A., {K{\'o}sp{\'a}l}, {\'A}., {{\'A}brah{\'a}m}, P., {et~al.}
  2015{\natexlab{b}}, \mnras, 447, 577

\bibitem[{{{\"O}berg} {et~al.}(2010){{\"O}berg}, {Qi}, {Fogel}, {Bergin},
  {Andrews}, {Espaillat}, {van Kempen}, {Wilner}, \& {Pascucci}}]{oberg2010}
{{\"O}berg}, K.~I., {Qi}, C., {Fogel}, J.~K.~J., {et~al.} 2010, \apj, 720, 480

\bibitem[{{Olofsson} {et~al.}(2016){Olofsson}, {Samland}, {Avenhaus},
  {Caceres}, {Henning}, {Mo{\'o}r}, {Milli}, {Canovas}, {Quanz}, {Schreiber},
  {Augereau}, {Bayo}, {Bazzon}, {Beuzit}, {Boccaletti}, {Buenzli}, {Casassus},
  {Chauvin}, {Dominik}, {Desidera}, {Feldt}, {Gratton}, {Janson}, {Lagrange},
  {Langlois}, {Lannier}, {Maire}, {Mesa}, {Pinte}, {Rouan}, {Salter},
  {Thalmann}, \& {Vigan}}]{olofsson2016}
{Olofsson}, J., {Samland}, M., {Avenhaus}, H., {et~al.} 2016, \aap, 591, A108

\bibitem[Ossenkopf \& Henning(1994)]{ossenkopf1994} Ossenkopf, V., \& Henning, T.\ 1994, \aap, 291, 943 

\bibitem[{{Pani{\'c}} {et~al.}(2013){Pani{\'c}}, {Holland}, {Wyatt}, {Kennedy},
  {Matthews}, {Lestrade}, {Sibthorpe}, {Greaves}, {Marshall}, {Phillips}, \&
  {Tottle}}]{panic2013}
{Pani{\'c}}, O., {Holland}, W.~S., {Wyatt}, M.~C., {et~al.} 2013, \mnras, 435,
  1037

\bibitem[Pawellek et al.(2014)]{pawellek2014} Pawellek, N., Krivov, A.~V., 
Marshall, J.~P., et al.\ 2014, \apj, 792, 65


\bibitem[Pecaut \& Mamajek(2016)]{pecaut2016} Pecaut, M.~J., \& Mamajek, E.~E.\ 2016, 
\mnras, 461, 794 


\bibitem[{{P{\'e}ricaud} {et~al.}(2017){P{\'e}ricaud}, {Di Folco}, {Dutrey},
  {Guilloteau}, \& {Pi{\'e}tu}}]{pericaud2017}
{P{\'e}ricaud}, J., {Di Folco}, E., {Dutrey}, A., {Guilloteau}, S., \&
  {Pi{\'e}tu}, V. 2017, \aap, 600, A62

\bibitem[{{Raman} {et~al.}(2006){Raman}, {Lisanti}, {Wilner}, {Qi}, \&
  {Hogerheijde}}]{raman2006}
{Raman}, A., {Lisanti}, M., {Wilner}, D.~J., {Qi}, C., \& {Hogerheijde}, M.
  2006, \aj, 131, 2290

\bibitem[{{Rhee} {et~al.}(2007){Rhee}, {Song}, {Zuckerman}, \&
  {McElwain}}]{rhee2007}
{Rhee}, J.~H., {Song}, I., {Zuckerman}, B., \& {McElwain}, M. 2007, \apj, 660,
  1556

\bibitem[{{Riviere-Marichalar} {et~al.}(2014){Riviere-Marichalar}, {Barrado},
  {Montesinos}, {Duch{\^e}ne}, {Bouy}, {Pinte}, {Menard}, {Donaldson}, {Eiroa},
  {Krivov}, {Kamp}, {Mendigut{\'{\i}}a}, {Dent}, \& {Lillo-Box}}]{rm2014}
{Riviere-Marichalar}, P., {Barrado}, D., {Montesinos}, B., {et~al.} 2014, \aap,
  565, A68

\bibitem[Roberge et al.(2000)]{roberge2000} Roberge, A., Feldman, P.~D., 
Lagrange, A.~M., et al.\ 2000, \apj, 538, 904

\bibitem[Roccatagliata et al.(2009)]{roccatagliata2009} Roccatagliata, V., Henning, T., Wolf, S., et al.\ 2009, \aap, 497, 409 

\bibitem[{{Rodigas} {et~al.}(2014){Rodigas}, {Debes}, {Hinz}, {Mamajek},
  {Pecaut}, {Currie}, {Bailey}, {Defrere}, {De Rosa}, {Hill}, {Leisenring},
  {Schneider}, {Skemer}, {Skrutskie}, {Vaitheeswaran}, \&
  {Ward-Duong}}]{rodigas2014}
{Rodigas}, T.~J., {Debes}, J.~H., {Hinz}, P.~M., {et~al.} 2014, \apj, 783, 21

\bibitem[{{Rosenfeld} {et~al.}(2013){Rosenfeld}, {Andrews}, {Hughes}, {Wilner},
  \& {Qi}}]{rosenfeld2013}
{Rosenfeld}, K.~A., {Andrews}, S.~M., {Hughes}, A.~M., {Wilner}, D.~J., \&
  {Qi}, C. 2013, \apj, 774, 16

\bibitem[{{Sheret} {et~al.}(2004){Sheret}, {Dent}, \& {Wyatt}}]{sheret2004}
{Sheret}, I., {Dent}, W.~R.~F., \& {Wyatt}, M.~C. 2004, \mnras, 348, 1282

\bibitem[{{Torres} {et~al.}(2008){Torres}, {Quast}, {Melo}, \&
  {Sterzik}}]{torres2008}
{Torres}, C.~A.~O., {Quast}, G.~R., {Melo}, C.~H.~F., \& {Sterzik}, M.~F. 2008,
  {Young Nearby Loose Associations}, ed. B.~{Reipurth}, 757

\bibitem[{{van Leeuwen}(2007)}]{vanleeuwen2007}
{van Leeuwen}, F. 2007, \aap, 474, 653

\bibitem[{{Vican} {et~al.}(2016){Vican}, {Schneider}, {Bryden}, {Melis},
  {Zuckerman}, {Rhee}, \& {Song}}]{vican2016}
{Vican}, L., {Schneider}, A., {Bryden}, G., {et~al.} 2016, \apj, 833, 263

\bibitem[Vidal-Madjar et al.(1994)]{vidal-madjar1994} Vidal-Madjar, A., 
Lagrange-Henri, A.-M., Feldman, P.~D., et al.\ 1994, \aap, 290, 245

\bibitem[Visser et al.(2009)]{visser2009} Visser, R., van Dishoeck, E.~F., \& Black, J.~H.\ 2009, \aap, 503, 323 

\bibitem[Webb et al.(1999)]{webb1999} Webb, R.~A., Zuckerman, B., Platais, I., et al.\ 1999, \apjl, 512, L63 


\bibitem[{{Williams} \& {Andrews}(2006)}]{williams2006}
{Williams}, J.~P., \& {Andrews}, S.~M. 2006, \apj, 653, 1480

\bibitem[{{Wilson} \& {Rood}(1994)}]{wilson1994}
{Wilson}, T.~L., \& {Rood}, R. 1994, \araa, 32, 191

\bibitem[{{Wyatt}(2008)}]{wyatt2008}
{Wyatt}, M.~C. 2008, \araa, 46, 339

\bibitem[{{Wyatt} {et~al.}(2015){Wyatt}, {Pani{\'c}}, {Kennedy}, \&
  {Matr{\`a}}}]{wyatt2015}
{Wyatt}, M.~C., {Pani{\'c}}, O., {Kennedy}, G.~M., \& {Matr{\`a}}, L. 2015,
  \apss, 357, 103

\bibitem[de Zeeuw et al.(1999)]{dezeeuw1999} de Zeeuw, P.~T., Hoogerwerf, R., de Bruijne, J.~H.~J., Brown, A.~G.~A., \& Blaauw, A.\ 1999, \aj, 117, 354

\bibitem[Zuckerman et al.(1995)]{zuckerman1995} Zuckerman, B., Forveille, T., \& Kastner, J.~H.\ 1995, \nat, 373, 494

\bibitem[Zuckerman \& Song(2012)]{zuckerman2012} Zuckerman, B., \& Song, I.\ 2012, \apj, 758, 77


\end{thebibliography}

\listofchanges

\end{document}